%% file: mn_mns_mp.tex
\documentclass[aps,prc,twocolumn,preprintnumbers,superscriptaddress,nofootinbib,tightenlines,floatfix]{revtex4-1}
\input{root_files/preamble}

\newcommand{\ccny}{
	Department of Physics, 
	The City College of New York, 
	New York, NY 10031, USA
	}
\newcommand{\graduatecenter}{
	Graduate School and University Center, 
	The City University of New York, 
	New York, NY 10016, USA
	}
\newcommand{\jlab}{
	Theory Group, Thomas Jefferson National Accelerator Facility, 
	Newport News, VA 23606, USA
	}
\newcommand{\lanl}{
	Theoretical Division, Los Alamos National Laboratory,
	Los Alamos, NM 87545, USA
	}
\newcommand{\lbl}{
	Nuclear Science Division, Lawrence Berkeley National Laboratory,
	Berkeley, CA 94720, USA
	}

\newcommand{\wm}{
	Department of Physics,
	The College of William \& Mary,
	Williamsburg, VA 23187, USA
	}

\newcolumntype{P}[1]{>{\centering\arraybackslash}p{#1}}

\begin{document}

{\count255=\time\divide\count255 by 60 \xdef\hourmin{\number\count255}
  \multiply\count255 by-60\advance\count255 by\time
  \xdef\hourmin{\hourmin:\ifnum\count255<10 0\fi\the\count255}}

\title{Strong isospin violation and 
chiral logarithms in the baryon spectrum}

\author{David Brantley}
\email[]{dabrantley01@email.wm.edu}
\affiliation{\footnotesize \wm}
\affiliation{\footnotesize \lbl}

\author{B\'{a}lint Jo\'{o}}
\email[]{bjoo@jlab.org}
\affiliation{\footnotesize \jlab}

\author{Ekaterina V. Mastropas}
\email[]{mastropaska@gmail.com}
\affiliation{\footnotesize \wm}
\affiliation{\footnotesize \jlab}

\author{Emanuele Mereghetti}
\email[]{emereghetti@lanl.gov}
\affiliation{\footnotesize \lanl}

\author{Henry~Monge-Camacho}
\email[]{hjmonge@email.wm.edu}
\affiliation{\footnotesize \wm}
\affiliation{\footnotesize \lbl}

\author{Brian Tiburzi}
\email[]{btiburzi@ccny.cuny.edu}
\affiliation{\footnotesize \ccny}
\affiliation{\footnotesize \graduatecenter}

\author{Andr\'{e} Walker-Loud}
\email[]{awalker-loud@lbl.gov}
\affiliation{\footnotesize \lbl}
\affiliation{\footnotesize \wm}
\affiliation{\footnotesize \jlab}



\begin{abstract}
We present a precise lattice QCD calculation of the contribution to the neutron-proton mass splitting arising from strong isospin breaking, 
$m_n-m_p|_{QCD}=2.32\pm0.17$~MeV.
We also determine $m_{\Xi^-} - m_{\Xi^0}|_{QCD} = 5.44\pm0.31$~MeV.
The calculation is performed at three values of the pion mass, with several values of the quark mass splitting and multiple lattice volumes, but only a single lattice spacing and an estimate of discretization errors.
The calculations are performed on the anisotropic clover-Wilson ensembles generated by the Hadron Spectrum Collaboration.
The omega-baryon mass is used to set the scale $a_t^{-1}=6111\pm127$~MeV, while the kaon masses are used to determine the value of the light-quark mass spitting.
The nucleon mass splitting is then determined as a function of the pion mass.
We observe, for the first time, conclusive evidence for non-analytic light quark mass dependence in lattice QCD calculations of the baryon spectrum.
When left as a free parameter, the fits prefer a nucleon axial coupling of $g_A=1.24(56)$.
To highlight the presence of this chiral logarithm in the nucleon mass splitting, we also compute the isospin splitting in the cascade-baryon system which is less sensitive to chiral dynamics.
Finally, we update the best lattice QCD determination of the CP-odd pion-nucleon coupling that would arise from a non-zero QCD theta-term, $\bar{g}_0 / (\sqrt{2}f_\pi) = (14.7\pm1.8\pm1.4) \cdot 10^{-3} \bar{\theta}$.

The original lattice QCD correlation functions, analysis results and extrapolated quantities are packaged in HDF5 files made publicly available including a simple Python script to access the numerical results, construct effective mass plots along with our analysis results, and perform the extrapolations of various quantities determined in this work.

\end{abstract}

\maketitle

\input{sections/1_introduction}

\input{sections/2_lqcd_details}

\input{sections/3_nucleon_splitting}

\input{sections/4_theta}

\input{sections/5_conclusions}

%
\acknowledgments
%

We would like to thank Kostas Orginos for many useful discussions and the use of the Sporades cluster at The College of William \& Mary, funded by MRI grant number PHY-0723103.
We would like to thank Chia Cheng Chang for extensive discussions regarding fitting systematics and Bayesian constrained fits.
We also thank Will Detmold for his involvement at early stages of the work.
The LQCD calculations for this work utilized the Chroma software suite~\cite{Edwards:2004sx} and the QUDA library~\cite{Clark:2009wm,Babich:2011np}.
The calculations were performed on the Sporades Cluster at The College of William and Mary, 
at the Jefferson Lab High Performance Computing Center on facilities of the USQCD Collaboration, which are funded by the Office of Science of the U.S. Department of Energy,
and on the Kraken Supercomputer using the Extreme Science and Engineering Discovery Environment (XSEDE)~\cite{xsede}, which is supported by National Science Foundation grant number ACI-1053575.
The $32^3\times256$ configurations were in part generated on
the Jaguar Cray XT5 System at Oak Ridge Leadership Computing 
Facility using allocations under the DOE INCITE program awarded to the USQCD 
Collaboration.  This research used resources of the Oak Ridge Leadership Computing Facility at the Oak Ridge National Laboratory, which is supported by the Office of Science of the U.S. Department of Energy under Contract No. DE-AC05-00OR22725.

This research was supported in part by;
The U.S. Department of Energy (DOE), Office of Science, Office of Workforce Development for Teachers and Scientists, Office of Science Graduate Student Research (SCGSR) program, administered by the Oak Ridge Institute for Science and Education for the DOE under contract number DE-SC0014664: DAB;
The U.S. DOE, Office of Science, Office of Nuclear Physics under contract DE-AC05-06OR23177, under which Jefferson Science Associates, LLC, manages and operates the Jefferson Lab: BJ, EVM, AWL;
The US DOE Office of Nuclear Physics and by the LDRD program at Los Alamos National Laboratory: EM;
The DOE Early Career Research Program, Office of Nuclear Physics under contract DE-SC0012180: EVM, AWL;
A joint City College of New York-RIKEN/Brookhaven Research Center fellowship, and by the U.S. National Science Foundation, under grant number PHY15-15738: BCT;
The U.S. Department of Energy, Office of Science: Office of Advanced Scientific Computing Research, Scientific Discovery through Advanced Computing (SciDAC) program under Award Number KB0301052: AWL; 
The Office of Nuclear Physics under Contract number DE-AC02-05CH11231: AWL,
The Office of Nuclear Physics Double-Beta Decay Topical Collaboration under Contract number DE-SC0015376: DAB, HMC, AWL;
The DOE Early Career Research Program, Office of Nuclear Physics under FWP Number NQCDAWL: DAB, HMC, AWL.

\bibliography{root_files/emc}

\end{document}

%% file: root_files/preamble.tex
\usepackage{graphicx}  
\usepackage{dcolumn}   
\usepackage{bm}        
\usepackage{amssymb}   
\usepackage{standalone}
\usepackage{enumitem}
\usepackage{xcolor}
\usepackage{slashed}
\usepackage{booktabs}
\usepackage{multirow}
\usepackage{amsmath}
\usepackage{stackrel}
\usepackage{rotating}
\usepackage{pifont}
\usepackage{mathtools}
\usepackage{hyperref}
\hypersetup{
    colorlinks=true,       
    linkcolor=blue,          
    citecolor=blue,        
    filecolor=blue,      
    urlcolor=blue           
}
\usepackage{simplewick}

\def\eqref#1{{(\ref{#1})}}
\def\mc#1{\mathcal{#1}}

\def\D{\Delta}
\def\L{\Lambda}

\def\S{\Sigma}

\def\O{\Omega}
\def\a{\alpha}
\def\b{\beta}
\def\d{\delta}
\def\e{\epsilon}
\def\g{\gamma}
\def\s{\sigma}

\def\ltap{\ \raise.3ex\hbox{$<$\kern-.75em\lower1ex\hbox{$\sim$}}\ }

%% file: sections/1_introduction.tex
\section{Introduction \label{sec:intro}}

Strong nuclear interactions exhibit a near perfect symmetry between protons and neutrons.
Today, 
we understand this symmetry as a manifestation of the approximate
$SU(2)$-flavor 
symmetry between the 
\emph{up}
and
\emph{down} 
quarks. 
Violation of 
$SU(2)$ 
symmetry is perturbatively small, 
but has profound consequences upon our understanding of the universe. 
Isospin breaking leads to a tiny 
relative splitting between the nucleon masses
($\sim0.07\%$), 
which allows for the neutron to undergo the weak $\b$-decay process. 
The primordial abundance of hydrogen and helium after big-bang nucleosynthesis is exquisitely sensitive to the magnitude of 
isospin breaking, due to the sensitivity of the weak-reaction rates of nucleons on the nucleon mass splitting. 
Varying the size of isospin breaking by only 1\%, 
for example, 
is inconsistent with the observed abundance of primordial nuclei at the two-sigma level~\cite{Banerjee:bbn}.
Explicit isospin breaking in QCD interactions arises due to the difference between up and down quark masses, 
and leads to charge symmetry breaking phenomena, 
see 
Ref.~\cite{Miller:2006tv} for an overview.
Isospin is additionally broken by the quark electric charges, 
and the Coulomb repulsion between protons has a significant influence on the nuclear landscape, 
from the stability of the Sun to neutron-rich exotic nuclei and fission.

Connecting isospin breaking in the Standard Model to that in nuclear physics is theoretically challenging due to the strongly coupled nature of low-energy QCD. 
Properties of strongly interacting matter, 
such as the nucleon mass splitting, 
can be computed using the non-perturbative numerical technique known as lattice QCD (LQCD).
LQCD utilizes the path-integral formulation of QCD on a discrete Euclidean spacetime lattice, 
and allows QCD correlation functions to be stochastically determined.
In order to make predictions directly from QCD, 
LQCD systematics must be sufficiently controlled.
Calculations must be performed at multiple lattice spacings, 
such that the continuum limit can be performed.
Multiple lattice volumes must be used to extrapolate to the infinite volume limit.
Finally, 
the input quark masses must be tuned and/or extrapolated to their physical values.
A comprehensive world-wide summary of LQCD calculations of important basic QCD quantities
can be found in the FLAG Working Group report~\cite{Aoki:2016frl}.
LQCD calculations utilize state-of-the-art high-performance computing, 
and the paramount goal of calculating properties of the lightest nuclei with all systematics controlled 
represents an \textit{exascale} challenge~\cite{np_exa}.
To maximize the impact of such non-trivial resource requirements,
the results need to be coupled to and understood within the broader field of nuclear physics.  
An essential tool to attain this is effective field theory (EFT)~\cite{Weinberg:1978kz}.  
The EFT description of low-energy QCD is chiral perturbation theory
($\chi$PT)%
~\cite{Langacker:1973hh,Gasser:1983yg,Leutwyler:1993iq}, 
which is formulated in terms of pion degrees of freedom as an expansion about the chiral limit.
EFTs are constructed by including all operators consistent with the symmetries of the theory.
While the form of these operators is dictated by symmetries, the values of the coefficients, 
known as low-energy constants (LECs), 
are \textit{a priori} unknown.
LECs must be determined by comparing derived formula to experimentally measured quantities, 
or by matching with numerical results from LQCD.
In particular, 
LQCD affords the ability to determine LECs of the quark-mass dependent operators, 
a feat which is considerably challenging or even impossible when comparing with experimental results alone.  
Prime examples can be found in 
Ref.~\cite{Aoki:2016frl}.
The true power of $\chi$PT is to economize on LQCD calculations, 
as the determination of LECs from QCD permits systematic EFT predictions for other quantities.

The efficacy of an EFT is determined by the size of its expansion parameters,  
which must be sufficiently small to organize contributions from the multitude of operators. 
The expansion parameter for two-flavor $\chi$PT is given by
$\varepsilon_{\pi} = m_\pi^2 / \Lambda_\chi^2$,
where $\Lambda_\chi\sim1$~GeV is a typical hadronic scale, and $m_\pi$ is the mass of the pion.
This small parameter provides for a rapidly converging expansion for pion masses up to a few hundred MeV~\cite{Durr:2013goa}.
If one also considers dynamical strange matter, 
then kaon and eta degrees of freedom are relevant.
The convergence of $SU(3)$ $\chi$PT is not as good as that of $SU(2)$ due, 
in part, 
to the number of virtual mesons propagating in loop diagrams,  
and the size of the expansion parameter
$\varepsilon_{K} = m_K^2 / \Lambda_\chi^2$.
Heavy matter fields, 
such as nucleons and mesons containing a heavy quark, 
can also be incorporated in the EFT.
This requires care, 
however, 
as the large mass scale can enter loop corrections and spoil the EFT power counting~\cite{Gasser:1987rb}.
A well-known solution to this problem is that of heavy baryon $\chi$PT (HB$\chi$PT)~\cite{Jenkins:1990jv,Jenkins:1991es},  
in which the theory is expanded about the infinite mass limit of the baryon, 
in a similar spirit to heavy-quark EFT~\cite{Georgi:1990um}.
More recently, an infrared regularization scheme was proposed~\cite{Becher:1999he} and extended to multi-loop integrals~\cite{Fuchs:2003qc,Schindler:2003xv} which treats the nucleon with a relativistic Lagrangian, but has a well-defined mapping onto the HB$\chi$PT power counting.
This new scheme automatically includes the proper kinematic singularities which appear in various quantities, 
whereas the HB$\chi$PT formulation requires summation of higher-order corrections for this to occur.

The hallmark prediction of $\chi$PT is the non-analytic behavior of quantities with respect to the pion mass. 
Often this behavior is logarithmic, and referred to generically as ``chiral logs." 
Because the square of the pion mass is proportional to the quark mass 
at leading order (LO) in $\chi$PT, 
these chiral logs are non-analytic in the light quark masses.
Such effects cannot be produced by a simple power-series expansion about the chiral limit, 
and are crucial predictions for QCD in the non-perturbative regime. 
Conclusive evidence for the predicted quark-mass dependence will expose 
\textit{chiral dynamics}
in low-energy QCD correlation functions, 
and further establish confidence in LQCD techniques. 
Such evidence has been observed in properties of light~\cite{Aoki:2008sm,Durr:2013goa} and heavy~\cite{Colquhoun:2015mfa} mesons and hadrons with a heavy (charm or bottom) quark~\cite{Detmold:2012ge}; but, 
to date, there has been no conclusive demonstration for such behavior in properties of baryons composed of light quarks.
Suggestive evidence for the presence of non-analytic light quark mass dependence in the baryon spectrum was presented in 
Ref.~\cite{WalkerLoud:2011ab} using various linear combinations of octet and decuplet baryon masses.
In this work, we present, for the first time, definitive evidence for a chiral log in the strong isospin splitting of the nucleon.
This evidence constitutes an important foundational step for broadening the reach of LQCD calculations when combined with $\chi$PT.

We begin with a description of the lattice calculation, 
our analysis strategies, 
and a determination of the lattice scale in 
Sec.~\ref{sec:lqcd}.
We use the kaon spectrum to determine the LQCD input value of $2\delta \equiv m_d -m_u$ that reproduces the physical value of the isovector light quark mass.
In Sec.~\ref{sec:nucleon}, 
we present our results for the isovector nucleon mass as a function of $m_d - m_u$ and $m_\pi$, and demonstrate the presence of non-analytic light quark mass dependence.
To solidify this observation, we also present results for the isovector cascade mass. 
As the cascade also forms an isospin doublet,  
$SU(2)$ $\chi$PT describing the cascade spectrum is largely identical in form to that of the nucleon, 
with only the numerical values of the LECs altered.
Phenomenologically, we know the cascade axial coupling is approximately 5 times smaller than that of the nucleon.
This in turn implies that the coefficient of the chiral logarithm is approximately 10 times weaker in the cascade splitting than the nucleon, 
which is observed in the numerical results.
We then briefly discuss the implications for the QCD $\theta$-term in Sec.~\ref{sec:theta}, 
before concluding in Sec.~\ref{sec:conclusions}.

%% file: sections/2_lqcd_details.tex
\section{Details of the Lattice Calculation \label{sec:lqcd}}

The calculations presented in this work are performed on the Hadron Spectrum Collaboration (HSC) anisotropic clover-Wilson ensembles~\cite{Edwards:2008ja,Lin:2008pr}. The HSC ensembles exist for a variety of light quark masses and volumes but just a single lattice spacing with fixed renormalized anisotropy $\xi = a_{s}/a_{t} = 3.5$.

We show the space-time dimensions in terms of lattice sites, and bare quark parameters of the datasets used in Tab.~\ref{tab:val_isospin}. The configurations were generated using an improved anisotropic gauge action comprised of combined plaquette and rectangle terms as described in \cite{Morningstar:1997ff}.  The action simulated two degenerate light quarks of mass $m_l$ and a single flavor for the strange quark with mass $m_s$, using the $O(a)$ improved Sheikoleslami Wohlert \cite{Sheikholeslami:1985ij} action also known as the Wilson-Clover, or simply ``Clover'' action. The fermion action also utilized so called ``Stout" smeared gauge links \cite{Morningstar:2003gk}. Two levels of smearing were performed, with a stout smearing weight of $\rho =0.22$.  The gauge link smearing was performed only in the spatial directions. The anisotropy parameters  and clover coefficients were tuned non-perturbatively, employing the Schr\"{o}dinger Functional method as discussed in~\cite{Edwards:2008ja}.

 The configurations were generated using the Hybrid Monte Carlo \cite{Duane1987216} algorithm, utilizing the Chroma code \cite{Edwards:2004sx}. The single-flavor strange quark term was simulated by Rational Hybrid Monte Carlo \cite{RHMC}.  A variety of  algorithmic tuning techniques were used to optimize the configuration generation process, including use of even-odd preconditioning, utilizing an anisotropic time-step in the molecular dynamics, splitting the molecular dynamics integration into several time-scales both in the sense  of mass preconditioning \cite{Urbach:2005ji} of the light quark determinant, and in the same spirit,  by splitting the gauge action into spatial and temporal parts and evolving the temporal gauge action with its larger forces on a finer timescale. Finally, the second order ``minimum norm" integrator of Omelyan \cite{Takaishi:2005tz,Omelyan2003272} was employed with an attempt to tune its parameter $\lambda$ to minimize the integration truncation errors.  The form of the actions and a majority of the gauge generation technique optimizations are described in detail in \cite{Lin:2008pr}, with the exception of some additional tuning for the larger lattices ( e.g. further tuning the integrator parameters ) that were carried out after that publication.

\begin{table*}
\begin{ruledtabular}
\begin{tabular}{cccccccccc}
\multicolumn{4}{c}{ensemble}& $m_\pi$& $m_K$& $N_{cfg}$& $N_{src}$& $a_t \d$& $a_t m_s^{val}$\\
$L/a_s$& $T/a_t$& $a_t m_l$& $a_t m_s$& MeV& MeV&\\
\hline
16& 128& -0.0830& -0.0743& 490& 629& 207& 16& \{0.0002,0.0004,0.0010\}& \{-0.0743,-0.0728,-0.0713\}\\
32& 256& -0.0840& -0.0743& 421& 588& 291& 10& \{0.0002,0.0004,0.0010\}& \{-0.0743,-0.0728,-0.0713\}\\
32& 256& -0.0860& -0.0743& 241& 506& 802& 10.5& \{0.0002\}& \{-0.0743,-0.0728,-0.0713\}\\
\end{tabular}
\end{ruledtabular}
\caption{\label{tab:val_isospin}
Summary of LQCD ensembles used in this work. 
We provide approximate values of the pion and kaon masses at the unitary point after our scale setting procedure~\ref{sec:scale}.  The number of configurations $N_{cfg}$ and average number of random sources $N_{src}$ as well as the values of the strong isospin breaking parameter $2\d=m_d-m_u$ used in the valence sector are provided.
In order to control the scale setting, we also use several partially quenched values of the valence strange-quark mass.}
\end{table*}

The two-point correlation functions of the ground-state hadrons are constructed for this work in a standard fashion.
We generate several gauge-invariant Gaussian-smeared sources~\cite{Frommer:1995ik} on each gauge configuration with random space-time locations~\cite{Beane:2009kya}.
From each source, we solve for the $light$ and $strange$ quark propagators.
For efficient solves, we utilize the deflated \texttt{eigcg} inverter~\cite{Stathopoulos:2007zi} on CPU machines and the \texttt{QUDA} library~\cite{Clark:2009wm} with multi-GPU support~\cite{Babich:2011np} on GPU enabled machines.
A point sink or gauge-invariant smeared sink is then applied to each quark propagator to construct PS (point-smeared) or SS (smeared-smeared) correlation functions.

In order to induce strong isospin breaking, we follow the suggestion of Ref.~\cite{WalkerLoud:2009nf} and spread the valence up and down quark masses symmetrically about the degenerate light quark mass
\begin{align}\label{eq:symm_isospin}
&m_u^{val} = m_l -\d\, ,&
&m_d^{val} = m_l +\d\, .
\end{align}
Because the valence and sea quark masses are not equal, this is a partially quenched (PQ) LQCD calculation with induced PQ systematics which must be removed through the use of PQ $\chi$PT~\cite{Bernard:1993sv,Sharpe:1997by,Sharpe:1999kj,Sharpe:2001fh,Chen:2001yi,Beane:2002vq,Sharpe:2003vy,Tiburzi:2005na}. 
In Ref.~\cite{WalkerLoud:2009nf}, it was shown in some detail this choice of \textit{symmetric isospin} breaking significantly suppresses the unitarity violating PQ effects.
Most importantly, it was demonstrated that the \textit{errors} from PQ do not enter isospin-odd quantities, such as $m_n-m_p$ until $\mc{O}(\d^3)$, well beyond the current precision of interest.
In Table~\ref{tab:val_isospin}, we list the ensembles used in this work, as well as the pion and kaon masses in MeV, as determined from our scale setting in Sec.~\ref{sec:scale}.
We further list the number of sources and the values of $a_t \d$ used in this work.
In order to fully control the scale setting, we also vary the valence strange quark mass $a_t m_s^{val}$.
We found the tuned value of $a_t m_s=-0.0743$ results in a strange quark mass slightly lighter than the physical one, see Sec.~\ref{sec:scale}.
The quality of the correlation functions we compute on these HSC ensembles are very good as can be inferred from the higher-statistics calculations on the same ensembles in Refs.~\cite{Lin:2008pr,Beane:2009kya}.
In this article, we only show the new isospin breaking results not presented previously.
However, the Python script and \texttt{hfd5} file accompanying this article can be used to generate the effective masses of all the correlation functions used in this work.

%
\subsection{Stochastic and systematic uncertainties of the ground-state spectrum}
%

In order to determine the stochastic and systematic uncertainties of the ground-state hadron spectrum, we employ a fitting strategy that is an evolution of that described in Ref.~\cite{Junnarkar:2013ac}.
Either multi-exponential (multi-cosh) fits or the Matrix Prony (MP) method is used to fit the baryon (meson) correlation functions~\cite{Beane:2009kya}.
A large set of reasonable choices of fit window, MP window, etc. are chosen and swept over, resulting in $\mc{O}(100)$ different fit choices for each correlation function.
Each fit is performed with a seeded bootstrap to preserve the correlations between various hadron correlation functions computed on the same ensembles, and to ensure the same number of stochastic results across all ensembles, resulting in $N_{bs}=500$ statistical samples for each quantity.
We also perform standard least-squares fits for all the fits in this systematic loop to assess the quality of each fit.
For each fit, a weight is constructed as
\begin{equation}
w_i = \frac{Q_i}{\sigma_{E,i}^2}
\end{equation}
where $\s_{E,i}$ is the stochastic uncertainty determined for the ground-state energy, and the quality of fit is defined
\begin{equation}
Q = \int_{\chi^2_{min}}^\infty d \chi^2 \mc{P}(\chi^2,d),
\end{equation}
with 
\begin{equation}
\mc{P}(\chi^2,d) = \frac{1}{2^{d/2} \Gamma \left(\frac{d}{2}\right)} 
	(\chi^2)^{\frac{d}{2}-1} e^{-\chi^2 / 2}
\end{equation}
being the probability distribution function for $\chi^2$ with $d$ degrees of freedom.
To assess the fitting systematic uncertainty, the fits from the systematic sweep are re-sampled with weight to generate $N_{sys}=500$ different systematic fits for each correlation function.  By re-sampling with weight, the resulting flat systematic distribution faithfully represents the weighted distribution of the original fits and allows us to enforce an equal number of systematic samples for every correlation function on every ensemble.

In order to properly preserve the correlations amongst various quantities computed on the same ensemble, for example the nucleon isospin splitting at different values of $a_t \d$, care must be taken to resample the systematic distributions in a correlated manner.
For example, the multi-exponential/MP fits for $m_n-m_p$ are aligned such that the choice of $t_{min}$, $t_{max}$, $n_{exp}$, etc. are the same for each value of $a_t \d$ on a given ensemble.  The weight factor is then taken as the average of the weights from each value of $a_t \d$, such that the seeded weighted re-sampling always chooses the fits from each $a_t \d$ in equal proportion, thus preserving the correlation between the samples.
If this careful alignment is not performed, the resulting $\chi^2$-minimum in the subsequent chiral extrapolations becomes at least an order of magnitude too small, as the correlations become ``washed out''.

The use of the full covariance matrix is critical in the subsequent analysis due to the correlations amongst results computed on the same sea-quark ensembles but with different values of the valence quark mass parameters.
The interested reader can repeat our analysis as both the original LQCD correlation functions, as well as our full analysis results are provided in accompanying \texttt{hdf5} files with some routines written in Python that can access the numerical results.

%
\subsection{Scale setting\label{sec:scale}}
%

\begin{table*}
\begin{ruledtabular} 
\begin{tabular}{cccccccccc}
V& $a_t m_l$& $a_t m_s$ & $a_t m_s^{val}$& $a_t m_{\pi^\pm}$& $a_t m_{K^\pm}$& $a_t m_{K^0}$& $a_t m_\O$& $l_\O$& $s_\O$ \\
\hline
$16^3\times128$& -0.0830& -0.0743& -0.0743& 0.0801(4)(1)& 0.1028(3)(1)& 0.1038(3)(1)& $0.301(3)(2)$& 0.0707(15)(9)& 0.1646(36)(22)  \\
$16^3\times128$& -0.0830& -0.0743& -0.0728& --& 0.1064(3)(1)& 0.1073(3)(1)& 0.307(3)(2)& 0.0680(14)(8)& 0.1739(35)(21) \\
$16^3\times128$& -0.0830& -0.0743& -0.0713& --& 0.1098(3)(1)& 0.1107(3)(1)& 0.313(3)(2)& 0.0656(12)(8)& 0.1825(35)(21) \\
$32^3\times256$& -0.0840& -0.0743& -0.0743& 0.0689(1)(2)& 0.0963(1)(1)& 0.0972(1)(1)& 0.293(2)(2)& 0.0553(7)(8)& 0.1627(19)(24) \\
$32^3\times256$& -0.0840& -0.0743& -0.0728& --& 0.1000(1)(1)& 0.1009(1)(1)& 0.299(2)(2)& 0.0530(6)(7)& 0.1722(19)(23) \\
$32^3\times256$& -0.0840& -0.0743& -0.0713& --& 0.1035(1)(1)& 0.1044(1)(1)& 0.305(2)(2)& 0.0509(5)(6)& 0.1810(19)(52) \\
$32^3\times256$& -0.0860& -0.0743& -0.0743& 0.0393(1)(1)& 0.08276(6)(7)& 0.08383(6)(6)& 0.275(1)(1)& 0.0205(2)(1)& 0.1629(13)(7)\\
$32^3\times256$& -0.0860& -0.0743& -0.0728& --& 0.08691(7)(5)& 0.08791(7)(5)& 0.282(1)(1)& 0.0195(1)(1)& 0.1725(13)(6) \\
$32^3\times256$& -0.0860& -0.0743& -0.0713& --& 0.09086(7)(5)& 0.09182(7)(5)& 0.289(1)(1)& 0.0186(1)(1)& 0.1816(13)(6) \\
\end{tabular}
\end{ruledtabular}
\caption{  \label{tab:m_ls_omega} Computed values of the hadron spectrum in lattice units and the corresponding values of $l_\O$ and $s_\O$.
These results are computed with the smallest value of $a_t\d=0.0002$.}
\end{table*}

In order to relate dimensionless quantities computed on the lattice to physical quantities comparable to experiment, a lattice scale must be determined.
There is ambiguity in choosing a scale-setting method, but all choices must 
result in the same continuum limit.
This ambiguity becomes more relevant when one has just a single lattice spacing, as in the present work.
We choose the omega baryon mass, 
$m_{\O}$, 
to set the scale in this work.
Using a hadronic scale allows for a direct comparison with experimental quantities, after electromagnetic corrections have been accounted for.
The omega baryon has mild light-quark mass dependence as it is composed of only strange valence quarks.
This also results in a rapidly convergent $SU(2)$ chiral extrapolation for $m_{\O}$~\cite{Tiburzi:2008bk}.

As the lattice ensembles were generated with a strange quark mass near its physical value, 
only a simple interpolation in the strange mass is needed.
In order to perform the necessary light and strange quark mass extrapolations to determine the scale, we utilize the two ratios of hadronic quantities~\cite{Lin:2008pr}
\begin{align}
	&l_{\O} \equiv \frac{m^{2}_{\pi}}{m^{2}_{\O}}\, ,&
	&s_{\O} \equiv \frac{2 m^{2}_{K} - m^{2}_{\pi}}{m^{2}_{\O}}\, .&
\end{align}
At LO (leading order) in $\chi$PT we have the relations
\begin{align}
&m_\pi^2 = 2 B m_l\, ,&
&m_K^2 = B(m_l + m_s)\, ,&
\end{align}
where we quote the isospin-averaged kaon mass.
In order to capture the strange-quark mass dependence, we compute the spectrum with 3 different values of the valence strange quark mass, with values provided in Table~\ref{tab:val_isospin}.
The omega baryon mass can then be determined for each choice of parameters and fit as a function of $l_\O$ and $s_\O$.
For the unitary points, one has the simple parameterization
\begin{equation} \label{eq:m_O}
	m_{\O} = m_{0} + c^{(1)}_{l} l_{\O} +  c^{(1)}_{s} s_{\O} + ...
\end{equation}
where the (...) denote terms higher order in $l_\O$ and $s_\O$.
We can use PQ$\chi$PT for the decuplet baryons%
~\cite{Tiburzi:2004rh}
to make an \textit{Ansatz} for the dependence on 
$a_t m^{val}_s$.
Using the LO $\chi$PT expressions for the meson masses, one has
\begin{equation} \label{eq:m_Opq}
	m_{\O}^{PQ} = m_{0}^{PQ} 
		+ c^{(1)}_{l} \left( l_{\O} + \frac{1}{2} s^{sea}_{\O} \right)
		+ c^{(1)}_{s} s^{val}_{\O}\, ,
\end{equation}
where 
$m_0 = m_0^{PQ} + \frac{1}{2}  c_l^{(1)} s^{sea}_\O$. 
Written in this way, 
the fit parameters  
$c^{(1)}_{l}$ 
and 
$c^{(1)}_{s}$ 
are found to agree between the unitary and PQ theories.

\begin{figure}
\includegraphics[width = \columnwidth]{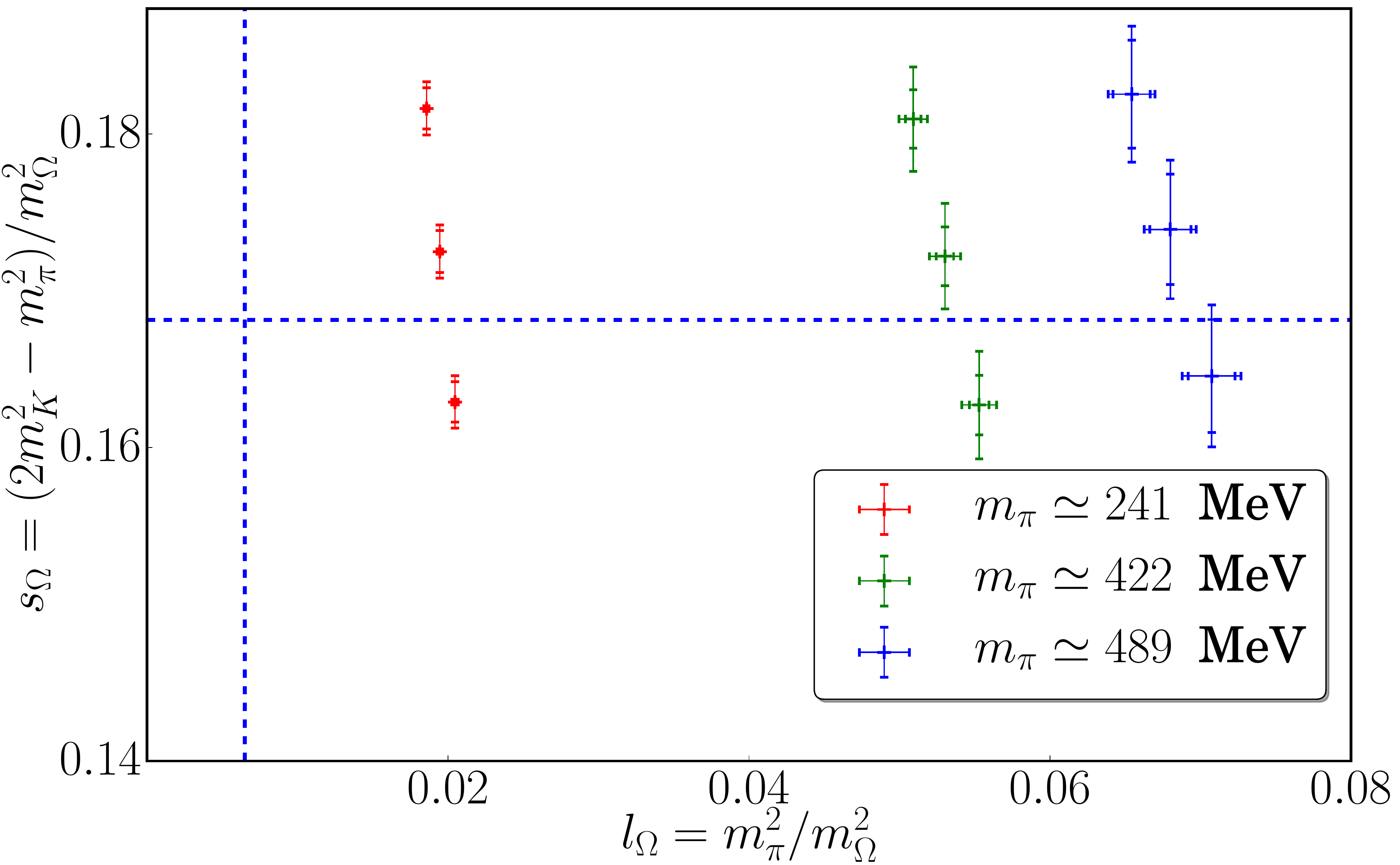}
\centering
\caption{\label{fig:l_omegaVs_omega} Parameter space of $l_{\O}$ and $s_{\O}$ used in this work.  The vertical and horizontal dashed lines denote the physical values of $l_\O$ and $s_\O$ with electromagnetic corrections subtracted.}
\end{figure}

The calculated values of $a_t m_\O$ are extrapolated to the physical point as functions of $l_\O$, $s^{val}_\O$ and $s^{sea}_\O$ using the above parameterizations.
Denoting quantities at the physical point with a $^*$ (e.g. $l^{*}_{\O} = m^{2 \, phys}_{\pi}  / m^{2 \, phys}_{\O}$), we can determine an ensemble independent scale
\begin{equation}
	a^{*}_{t} \equiv \frac{a_{t} m_{\O}(l^{*}_{\O},s^{*}_{\O})}{m^{phys}_{\O}}\, .
\end{equation}
The parameter space of $l_\O$ and $s_\O$ used in this work is depicted in Figure~\ref{fig:l_omegaVs_omega} and listed in Table~\ref{tab:m_ls_omega}.

In this work, we are interested in the strong isospin-breaking corrections to $m_n-m_p$.
We therefore define the physical point in the absence of electromagnetic corrections.
Unless the electromagnetic corrections to $m_\O$ are unnaturally large (greater than several MeV), these corrections are sub-percent and well within our total uncertainty budget, and therefore we choose to neglect them.
The strong isospin breaking in the pion spectrum is $\mc{O}(\d^2)$ and assumed to be small~\cite{Aoki:2016frl}.  
Further, the electromagnetic corrections to the $\pi^0$ are suppressed~\cite{Bijnens:1996kk}.
We therefore take the QCD value of $m_\pi$ in the absence of electromagnetism as defined by $m_{\pi^0}$.
The FLAG~\cite{Aoki:2016frl} estimate of the electromagnetic self-energy corrections to the kaon spectrum can be used to define the QCD value of the isospin-averaged kaon mass $m_K^{QCD}=494$~MeV 
(the uncertainties on these QCD input values are well within our total uncertainty).
The physical point is then defined in this work as $m_\O^{phys} \equiv m_{\O}^{PDG}$ and
\begin{align}
	l_\O^* &\equiv \frac{m_{\pi^0}^2}{m_{\O}^{2,PDG}} = 0.0065\, ,
\nonumber\\
	s_\O^* &\equiv \frac{2m_{K}^{2,QCD}-m_{\pi^0}^2}{m_{\O}^{2,PDG}} = 0.1681\, .
\end{align}

\begin{figure}
\begin{tabular}{c}
\includegraphics[width = \columnwidth]{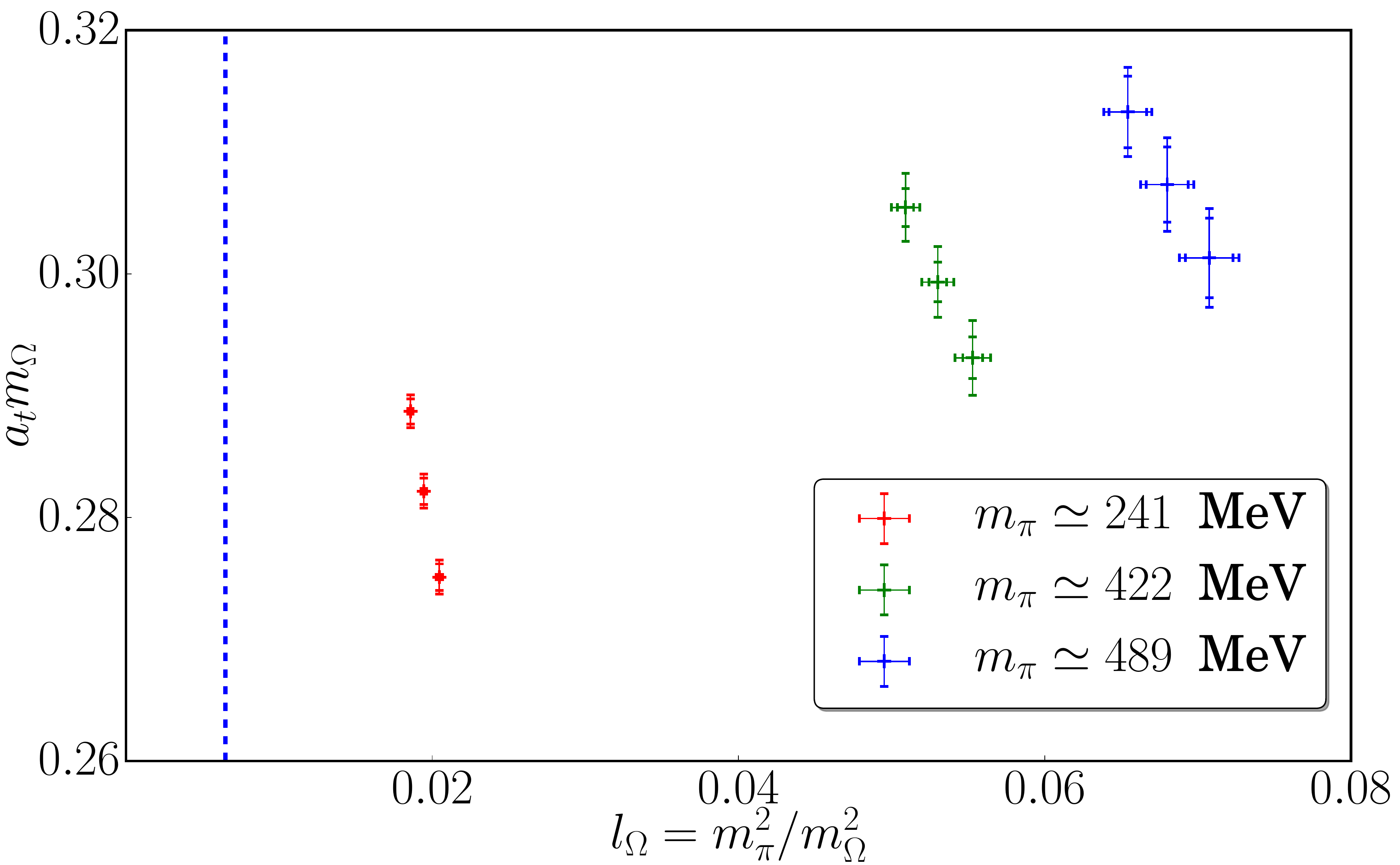} 
\end{tabular}
\caption{\label{fig:mO_lO} The $l_\O$ dependence of $a_{t}m_{\O}$.  The dashed vertical lines denote the physical value of $l_\O^*$.}
\end{figure}

\begin{figure}
\begin{tabular}{c}
\includegraphics[width = \columnwidth]{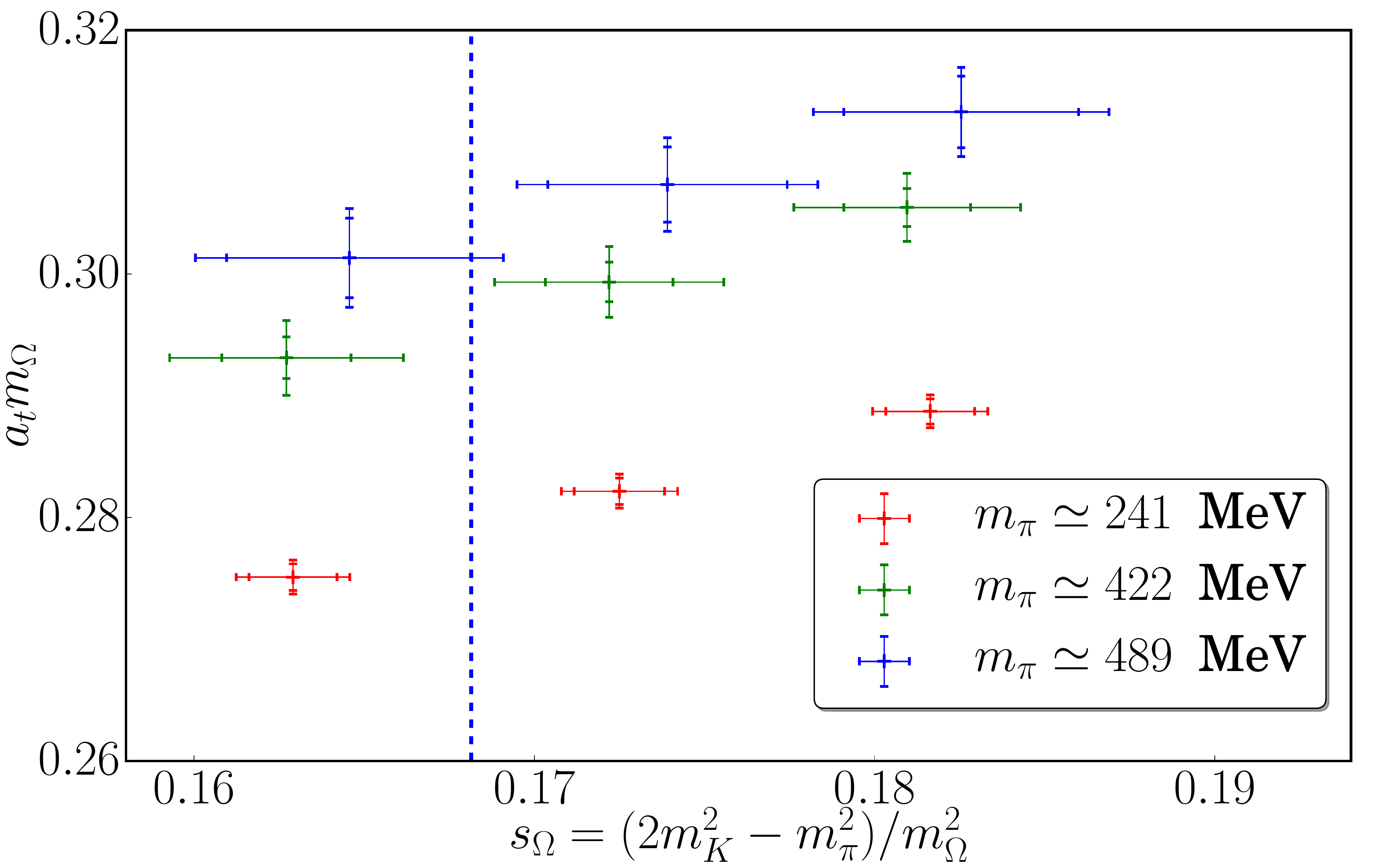}
\end{tabular}
\caption{\label{fig:mO_sO} 
The $s_\O$ dependence of $a_{t}m_{\O}$.  The dashed vertical lines denote the physical value of $s_\O^*$.}
\end{figure}

\begin{table*}
\begin{ruledtabular}
\begin{tabular}{c|ccc|ccc|ccl}
PQ& $a_t m_0$ & $c_l^{(1)}$& $c_s^{(1)}$
& $\chi^2$& dof& $Q$ & $a_t m_\O^{phys}$& $a_t$[fm]& $a_t^{-1}$[MeV]
\\
\hline
no& 
0.139(07)(04)& 0.50(13)(8)& 0.77(2)(01)& 0.39 & 4& 0.98
& 0.2721(35)(18)& 0.0320(4)(2)& 6145(80)(40)\\
yes& 
0.099(19)(23)& 0.50(14)(7)& 0.77(3)(11)& 0.38 & 4& 0.98
& 0.2736(38)(56)& 0.0322(4)(7)& 6111(85)(94)
\end{tabular}
\end{ruledtabular}
\caption{\label{tab:mO_v_lO_sO} 
Scale-setting extrapolation using Eqs.~\eqref{eq:m_O} and \eqref{eq:m_Opq}.
}
\end{table*}

In Figures~\ref{fig:mO_lO} and \ref{fig:mO_sO}, we depict the values of $a_t m_\O$ vs. $l_\O$ and $s_\O$.
One observes linear dependence of $a_t m_\O$ in both $l_\O$ and $s_\O$, consistent with the LO expressions in Eqs.~\eqref{eq:m_O} and \eqref{eq:m_Opq}.
Using the results listed in Table~\ref{tab:m_ls_omega}, the omega mass is determined as a function of $l_\O$ and $s_\O$.
Both the unitary and PQ formula fit the data well with the following caveat: 
a fully correlated fit to all data, including errors in the independent variables, produces an unexpectedly large $\chi^{2}$ despite having small residuals, normalized by the extrapolated uncertainty.
Removing the two heaviest valence strange quark masses, $a_t m_s = \{-0.0728, -0.0713\}$, in the two heaviest sea ensembles from the fit produces a much better  $\chi^{2}$, while losing none of the predictive power of the fit, even for the points not included. Using these quantities for the fit, 
we tabulate our fit results in Table \ref{tab:mO_v_lO_sO}.
We find the unitary and PQ fits results are perfectly consistent, with the PQ fit having a factor of 2 larger systematic uncertainty.
We take the PQ fit as our determination of the scale:
\begin{equation}\label{eq:physical_scale}
	\frac{1}{a^{*}_{t}} = 6111 \pm 85 \pm 94 \enspace \textrm{MeV},
\end{equation}
where the first uncertainty is statistical and the second is systematic.
The statistical and systematic uncertainties can be individually determined by taking the complete statistical-systematic covariance matrix constructed from the $N_{bs}\times N_{sys}$ samples, and first averaging over the systematic or statistical fluctuations, respectively.

\subsection{The kaon spectrum and determination of $\delta$}

In order to determine the QCD contribution to the nucleon isovector mass, we must first determine the physical value of $\d$.
At LO in $\chi$PT, the kaon masses are
\begin{align}
&m_{K^\pm}^2 = B (m_s + m_u)\, ,&
&m_{K^0}^2 = B (m_s + m_d)\, .&
\end{align}
A calculation of $\D m_K^2 \equiv m_{K^0}^2 - m_{K^\pm}^2 = 2B\d$ allows for this determination.
The electromagnetic contributions to this kaon splitting must be subtracted.
We use the value of the strong isospin splitting provided in the FLAG report
\begin{equation}\label{eq:ksplit_flag}
\Delta m^{2}_{K}\Big|_{QCD} = 5930 \textrm{ MeV}^2\, .
\end{equation}

\begin{table}
\begin{ruledtabular}
\begin{tabular}{ccccc}
$a_t m_l$& $a_t m_s$& $m_\pi$ [MeV]& $a_t \d$& $(a_t \D m_K)^2$\\
\hline
-0.0860& -0.0743& 241& 0.0002& 0.000178(03) \\
-0.0840& -0.0743& 421& 0.0002& 0.000189(02) \\
-0.0830& -0.0743& 490& 0.0002& 0.000196(05) \\
-0.0840& -0.0743& 421& 0.0004& 0.000378(03) \\
-0.0830& -0.0743& 490& 0.0004& 0.000392(06) \\
-0.0840& -0.0743& 421& 0.0010& 0.000947(05) \\
-0.0830& -0.0743& 490& 0.0010& 0.000980(12) \\
\end{tabular}
\end{ruledtabular}
\caption{\label{kaonSplittingTable}
Kaon mass splitting versus $\d$ and $m_\pi$ on the various ensembles. }
\end{table}

The values of the kaon mass splitting computed in this work are provided in Table~\ref{kaonSplittingTable}.  The kaon splitting exhibits a slight pion-mass dependence, indicating the presence of NLO (next-to-leading order) corrections.
We do not observe any 
$a_t m_s^{val}$ dependence.
Starting from the work of Gasser and Leutwyler~\cite{Gasser:1984gg}, we can integrate out the strange quark contribution to $\D m_K^2$ to arrive at the NLO formula 
\begin{equation} \label{eq:kaonSplitting}
	\D m^{2}_{K} = 2B \d \left[ 1 
		+ \a(\mu) m^{2}_{\pi} 
		- \frac{m^{2}_\pi}{(4 \pi f)^{2}} \ln \left(  \frac{m^{2}_{\pi}}{\mu^{2}}\right)
	\right]\, .
\end{equation}
In this expression, $\a(\mu)$ is an unknown LEC and $f$ is the pion decay constant in the chiral limit with the normalization $f_\pi=130$~MeV and we set $\mu=770$~MeV.%
\footnote{In this and all subsequent $\chi$PT analyses, we work at the standard $\chi$PT renormalization scale $\mu = 770$~MeV and treat an $\mu$-dependence of LECs implicitly.} 
In our $\chi$PT analysis, we use the FLAG $N_{f} = 2 + 1$ determination of $f=122.6$~MeV as input to the fits.
This expression describes the data well as observed in our analysis results collected in Table~\ref{tab:deltaFit}.
\begin{table}
\begin{ruledtabular}
\begin{tabular}{cc|cc|cl}
 $a_{t} B$ & $\a$
& $\chi^2/$dof& $Q$ & $a_{t}\delta^{*} \mathrm{[10^{-4}]}$
\\
\hline 
 0.411(6)(5)& 13.5(3)(3)& 2.25/5& 0.81 &1.87(2)(2)\\
\end{tabular}
\end{ruledtabular}
\caption{\label{tab:deltaFit} 
Extrapolation of $\D m_K^2$ using Eq.~\eqref{eq:kaonSplitting} and the determination of $a_t\d^*$.  Notice, we treat the scale dependence of the LEC $\a$ implicitly, because we work at the standard renormalization scale $\mu=770$~MeV throughout.
}
\end{table}

\begin{figure}
\includegraphics[width = \columnwidth]{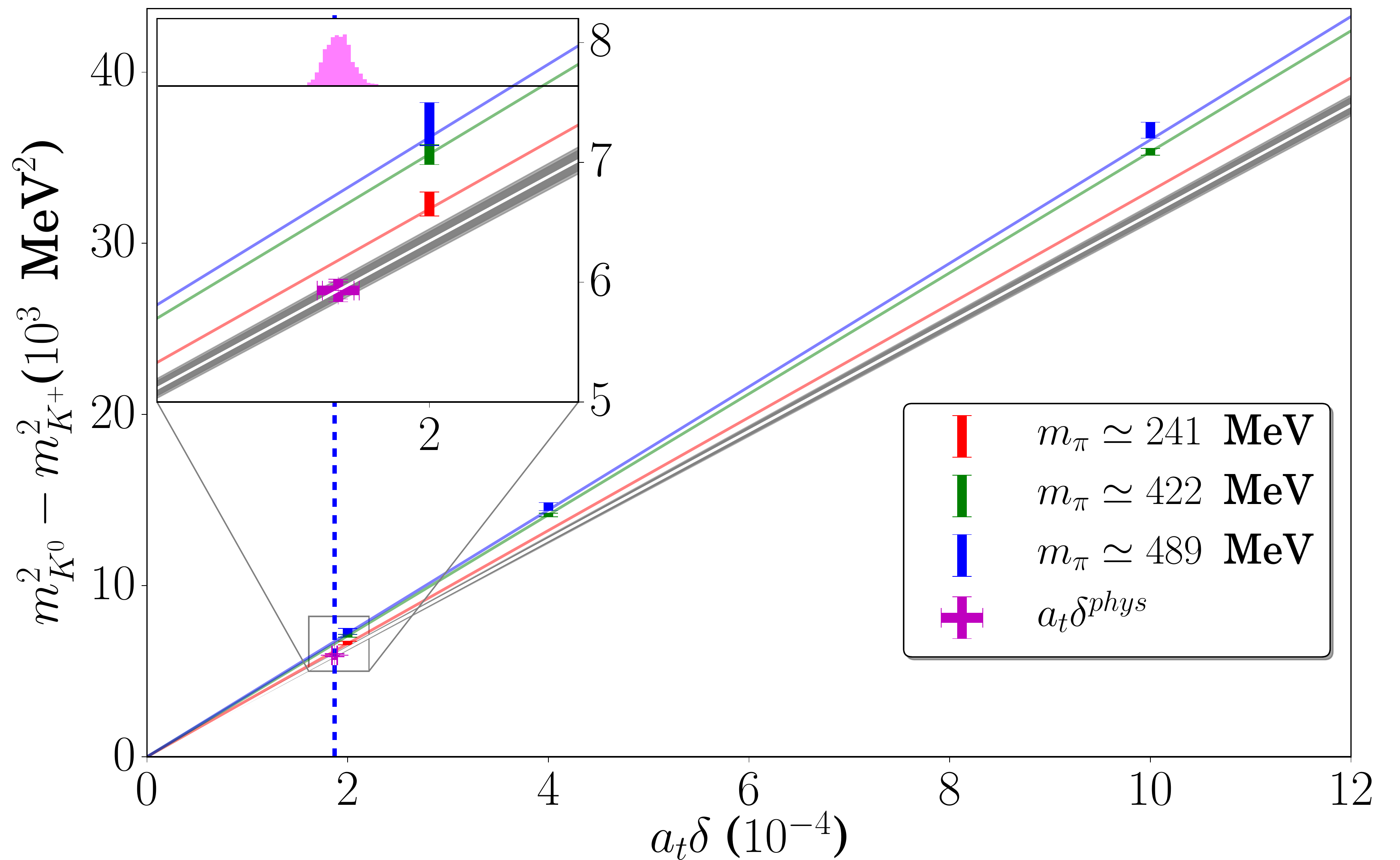}
\centering
\caption{\label{fig:ksplit}
Plot of strong isospin splitting $\Delta m^{2}_{K}$ versus lattice values for $\d$. 
The gray band is the predicted value of $\D m_K^2(\d,m_\pi=m_\pi^{phy})$ while the colored lines are the central values for the corresponding pion masses. The magenta point is the distribution of $a_{t} \d^{phys}$. }
\end{figure}

We solve for the value of $a_t \d$ that reproduces the physical QCD kaon splitting, Eq.~\eqref{eq:ksplit_flag}, finding 
\begin{equation} \label{eq:delta}
a_{t} \d^{*}  = 1.87(2)(2) \times 10^{-4}\, ,
\end{equation}
where the first and second uncertainties arise from the stochastic and systematic uncertainties, 
respectively.
The extrapolation and determination of $a_t\d^*$ are depicted in Figure~\ref{fig:ksplit}.
Interestingly, we can use the value of $a_t B$ determined in this fit to estimate the bare vacuum condensate, 
$\S = B F^2$, with $F=86.6$~MeV as the pion decay constant in the $F_\pi=92.2$~MeV normalization.  Using our lattice scale, Eq.~\eqref{eq:physical_scale}, we obtain the bare value
\begin{equation}\label{eq:bareSigma}
	\mathring{\Sigma}^{1/3} = 266(4)(1) \textrm{ MeV}\, ,
\end{equation}
which is very similar to the FLAG average~\cite{Aoki:2016frl}.
As we have not performed the necessary renormalization, 
this is a qualitative comparison.
However, it seems to imply the isovector quark-mass renormalization is close to unity.

%% file: sections/3_nucleon_splitting.tex
\section{Isovector Nucleon Mass and Chiral Logarithms\label{sec:nucleon}}

\begin{figure}
\begin{tabular}{c}
\includegraphics[width=\columnwidth]{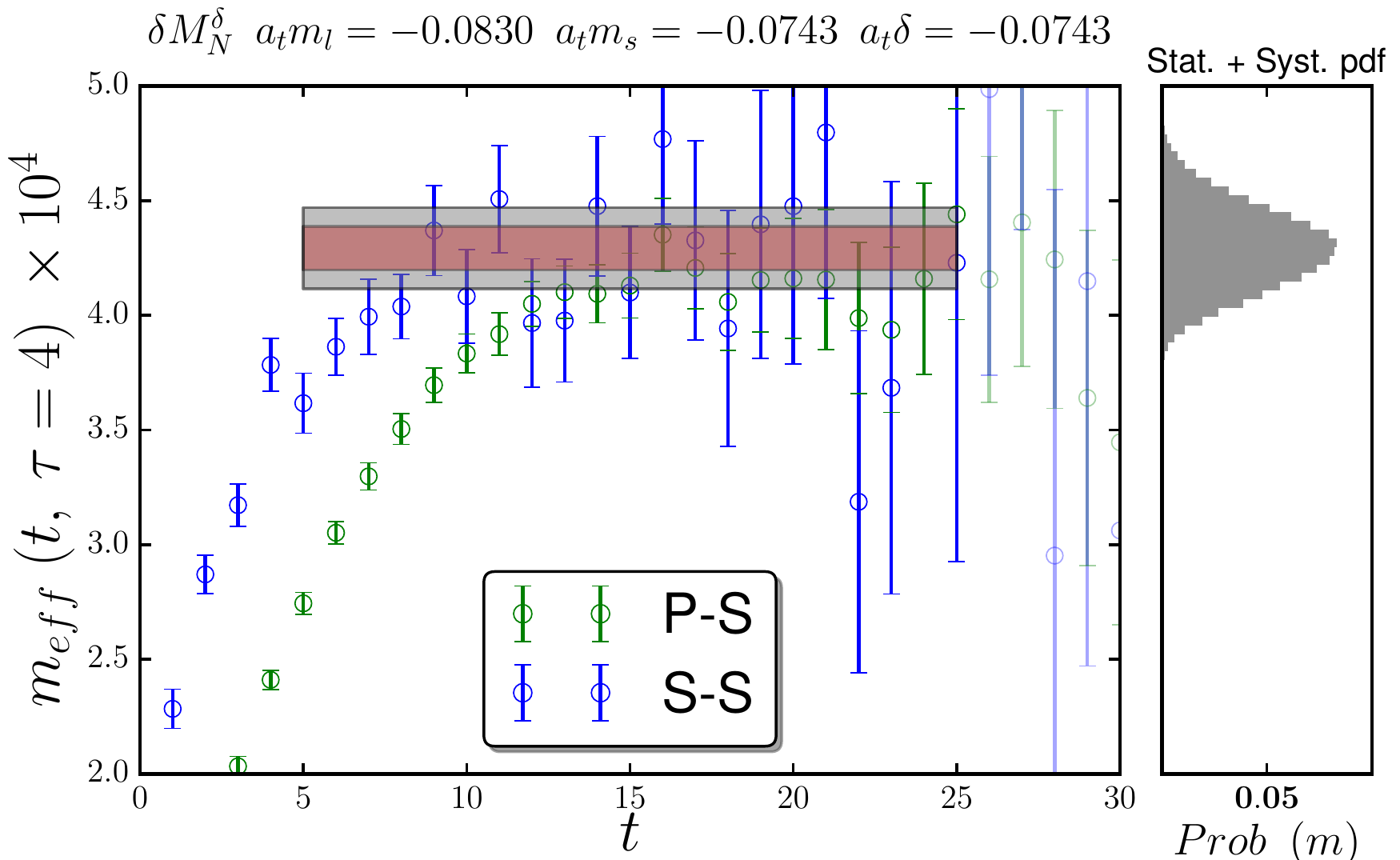}
\\
\includegraphics[width=\columnwidth]{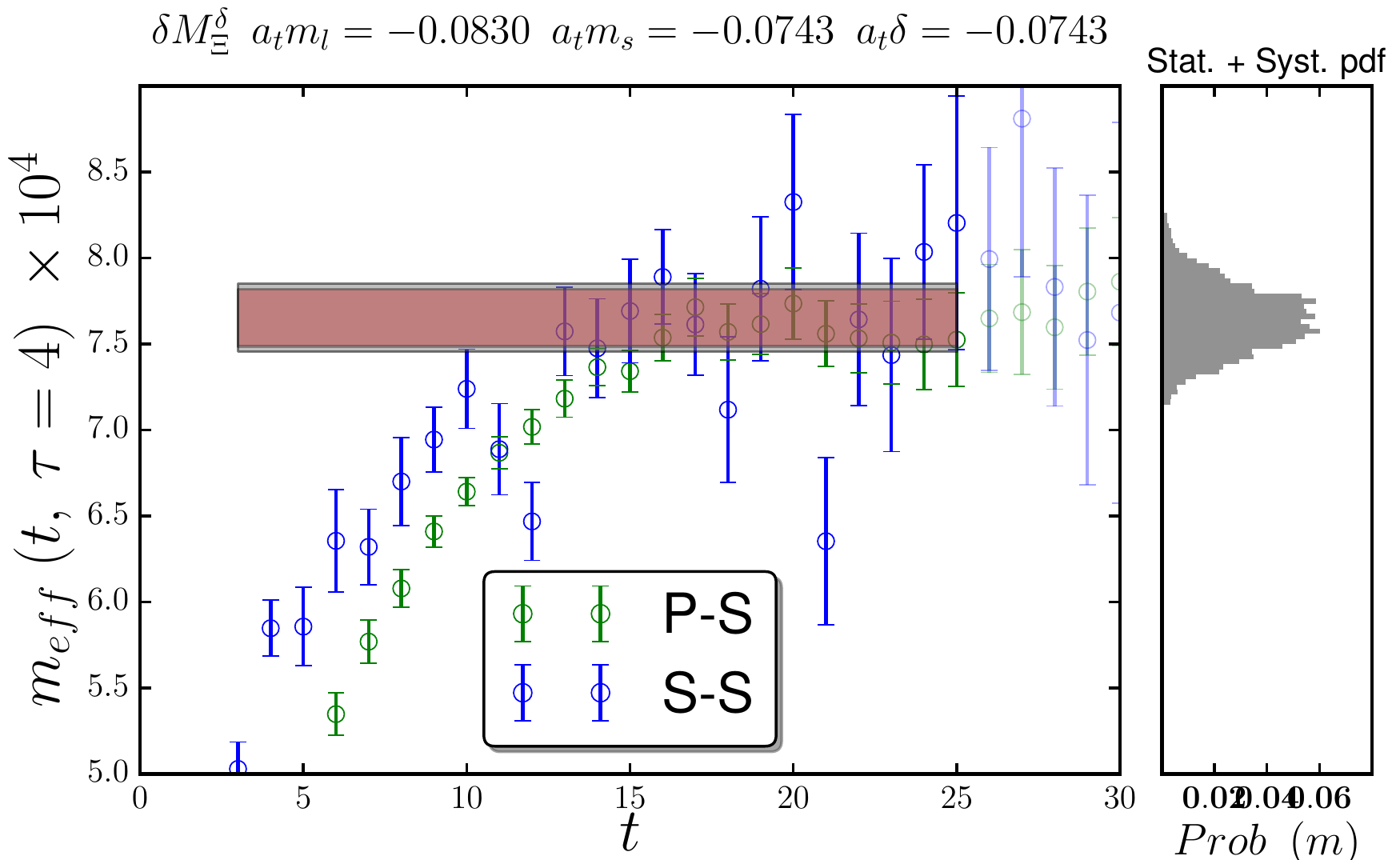}
\\
\includegraphics[width=\columnwidth]{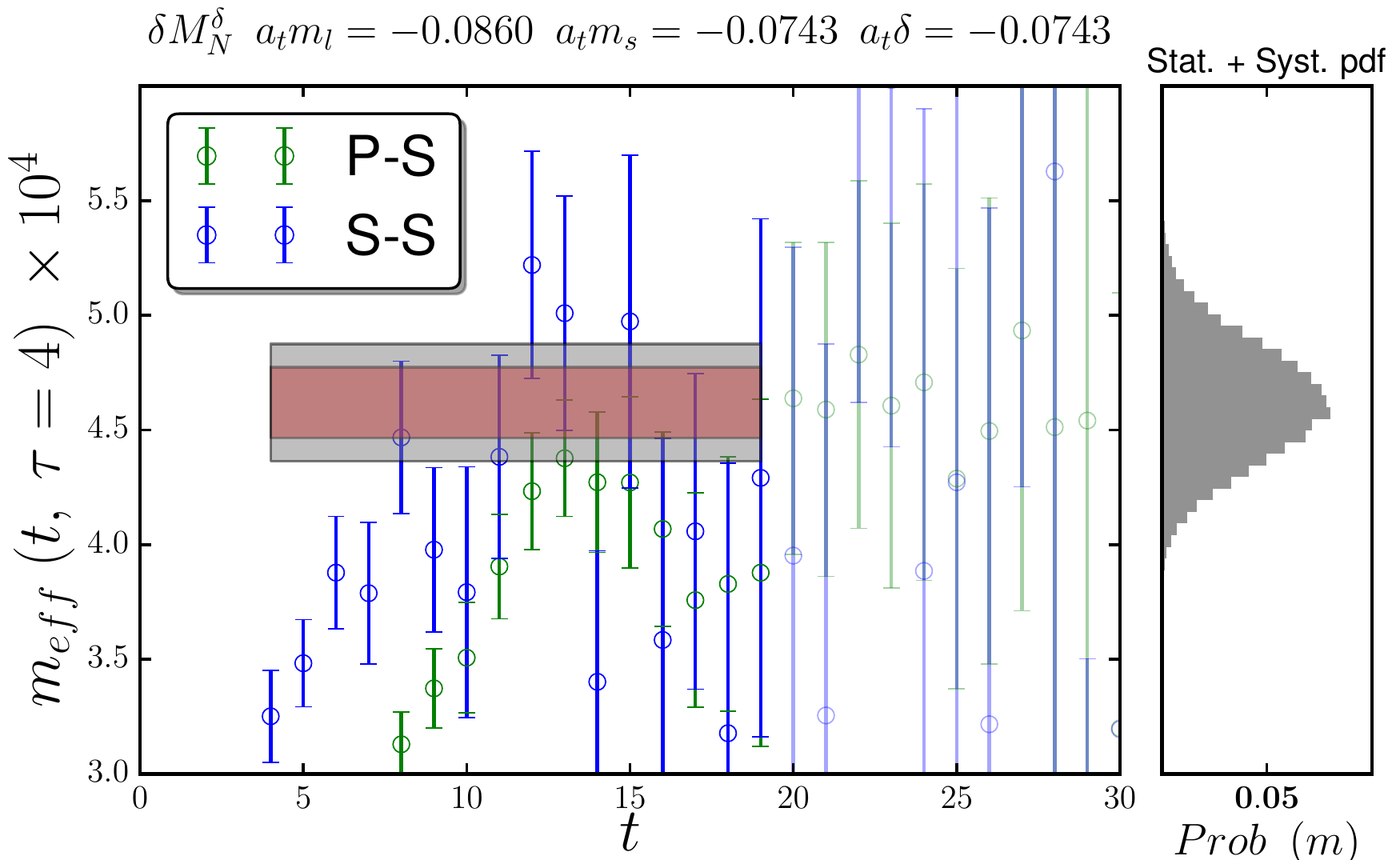}
\\
\includegraphics[width=\columnwidth]{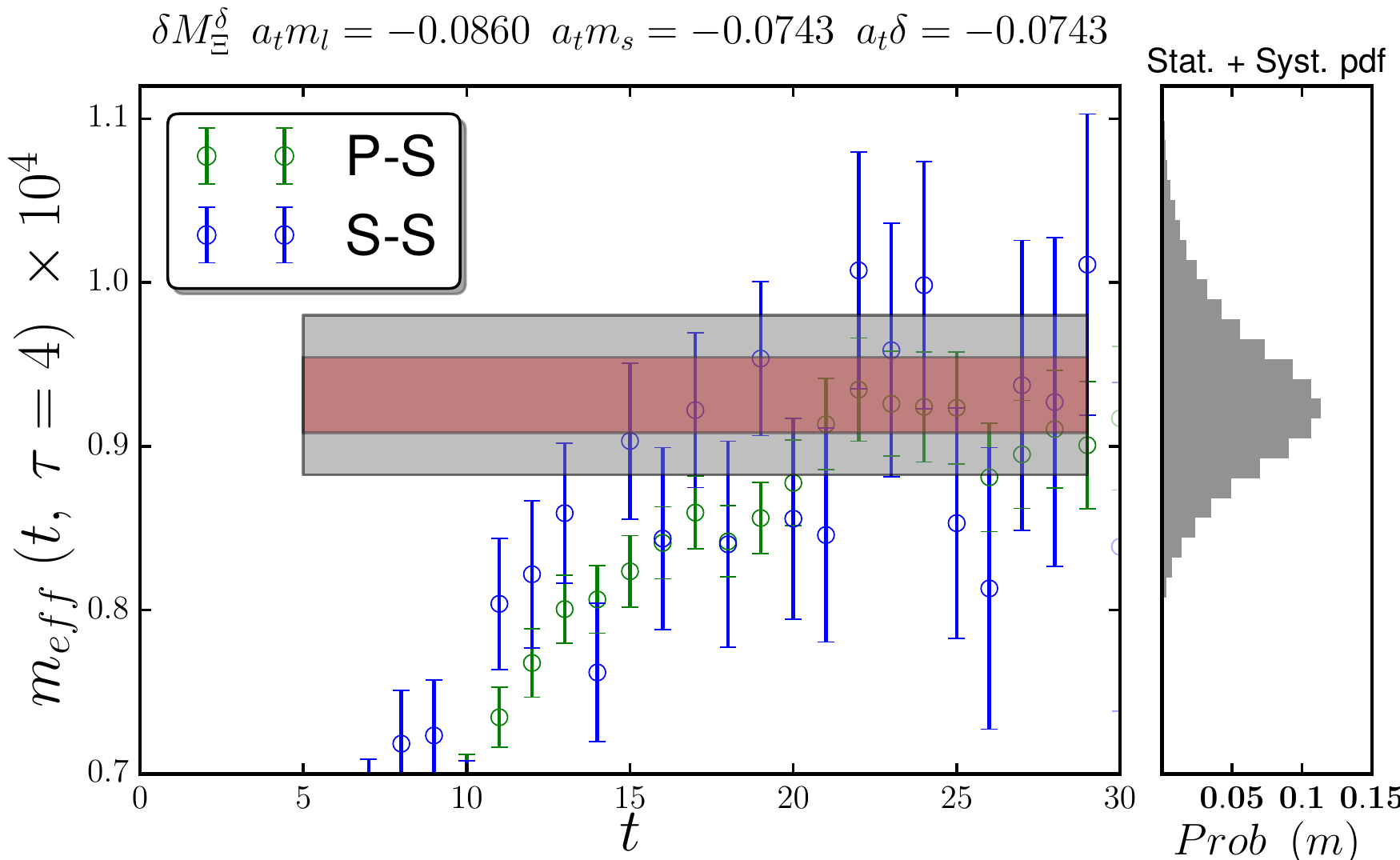}
\end{tabular}
\caption{\label{fig:eff_mass} 
Sample effective masses of the nucleon and cascade isovector correlation functions.
The resulting ground-state mass splitting determined from Matrix Prony and multi-exponential fits is displayed as a horizontal band over the region of times considered.}
\end{figure}

\begin{table}
\begin{ruledtabular}
\begin{tabular}{cccccc}
$a_t m_l$& $a_t m_s$& $m_\pi/\textrm{MeV}$& $a_t \d$& $a_t \d M_N^\d / (a_t\d)$& $a_t \d M_\Xi^\d / (a_t\d)$ \\
\hline
-0.0860& -0.0743& 241& 0.0002& 2.31(08)(09)& 4.66(12)(19) \\
-0.0840& -0.0743& 421& 0.0002& 2.34(04)(05)& 4.29(05)(10) \\
-0.0840& -0.0743& 421& 0.0004& 2.34(04)(05)& 4.31(05)(10) \\
-0.0840& -0.0743& 421& 0.0010& 2.33(04)(05)& 4.30(05)(10) \\
-0.0830& -0.0743& 490& 0.0002& 2.15(05)(07)& 3.83(08)(04) \\
-0.0830& -0.0743& 490& 0.0004& 2.15(05)(07)& 3.83(08)(04) \\
-0.0830& -0.0743& 490& 0.0010& 2.14(05)(07)& 3.83(08)(04) \\
\end{tabular}
\end{ruledtabular}
\caption{\label{tab:mn_split}
The nucleon ($a_t \d M_N^\d$) and cascade ($a_t \d M_\Xi^\d$) mass splittings, normalized by $a_t\d$ for different values of $a_t\d$ and $m_\pi$ on the various ensembles.}
\end{table}

We now turn to the nucleon mass splitting.
We define the isovector masses to be the positive quantities
\begin{align}
& \d M_N^\d \equiv m_n - m_p\, ,&
& \d M_\Xi^\d \equiv m_{\Xi^-} - m_{\Xi^0}\, .&
\end{align}
In Table~\ref{tab:mn_split}, we list the numerical values of the nucleon and cascade mass splittings determined in this work.
If Figure~\ref{fig:eff_mass}, we display sample effective mass plots of the nucleon and cascade isovector masses.
These values can be converted to MeV using the scale Eq.~\eqref{eq:physical_scale} and the physical value of $a_t \d^*$,
Eq.~\eqref{eq:delta}.

At LO in the chiral expansion, we write the heavy baryon Lagrangian~\cite{Jenkins:1990jv} with the conventions of Ref.~\cite{Tiburzi:2005na}, 
with the replacement $\a_N = -4 \a_M$ of that work, 
\begin{align}\label{eq:LO_N}
\mc{L}_N^{(LO)} =&\ \bar{N} i v \cdot D N 
	-\bar{T}^\mu iv \cdot D T_\mu +\D \bar{T}^\mu T_\mu
\nonumber\\&
	-\frac{\a_N}{2} \bar{N} \mc{M} N
	 +2\g_M \bar{T}^\mu \mc{M} T_\mu 
\end{align}
where $\mc{M} = \frac{1}{2} (\xi^\dagger m_q \xi^\dagger + \xi m_q^\dagger \xi)$.
We similarly construct the $SU(2)$ Lagrangian for the $\Xi,\Xi^*$ system, following Ref.~\cite{Tiburzi:2008bk}, but keep the normalization similar to that in Ref.~\cite{Tiburzi:2005na} instead of using the extra $1/(4\pi f)$ in the LO operators,
\begin{align}\label{eq:LO_Xi}
\mc{L}_\Xi^{(LO)} =&\ \bar{\Xi} i v \cdot D \Xi 
	-\bar{\Xi}^{*,\mu} iv \cdot D \Xi^*_\mu +\D_{\Xi^* \Xi} \bar{\Xi}^{*,\mu} \Xi^*_\mu
\nonumber\\&
	-\frac{\a_\Xi}{2}\, \bar{\Xi} \mc{M} \Xi 
	 -2\a_{\Xi^*} \bar{\Xi}^{*,\mu} \mc{M} \Xi^*_\mu
\end{align}

The present choice of normalization is such that the LO isovector masses are proportional to $\d$,
with the slopes $\d M_{N,\Xi}^\d = \a_{N,\Xi} \d$.
The NLO contributions that scale as $m_\pi^3$ for the isoscalar mass exactly cancel in the isovector mass, 
provided one utilizes the symmetric PQ isospin breaking, Eq.~\eqref{eq:symm_isospin}, or includes isospin breaking in the sea quarks with a unitary calculation.
The first non-vanishing corrections arise at NNLO (next-to-next-to-leading order), originating from the self-energy corrections due to virtual pion loops.
These long-range corrections depend logarithmically on the pion mass;
and, 
provided they have a large coefficient, 
cannot be well parameterized by a low-order power-series expansion about the chiral limit.
It is precisely this non-analytic behavior that signals the influence of chiral dynamics in QCD observables.

At NNLO in the $SU(2)$ chiral expansion, the expression for the nucleon mass splitting, including partial quenching effects, is given by~\cite{WalkerLoud:2009nf}
\begin{multline}\label{eq:dMN_full}
\delta M_{N}^{\delta} = \delta \bigg\{
	\a_N \left[1
		-(6g_A^2+1)\frac{m_\pi^2}{(4\pi f_{\pi})^2}\ln \left(\frac{m_{\pi}^2}{\mu^2}\right)
	\right] 
\\
	+ 4 g_{\pi N\D}^2 \left(\frac{20}{9} \g_M -\a_N \right) \frac{\mc{J}(m_\pi,\D,\mu)}{(4\pi f_\pi)^2}
\\
	+ \beta(\mu)\frac{2m_{_\pi}^2}{(4 \pi f_{\pi})^2}
	+\frac{\a_N \D_{PQ}^4}{2 m_\pi^2 (4\pi f_\pi)^2}(4 - 3g_0^2)
	\bigg\}\, .
\end{multline}
In this expression, all finite contributions are absorbed into the LECs which stem from local operators.
The quantity 
$\a_N\d$ is the LO contribution to $\d M_N^\d$ and similarly, the LO contribution to the delta-resonance isospin splitting is proportional to $\g_M\d$, e.g.
\begin{equation}\label{eq:gamma_m}
m_{\D^+} - m_{\D^{++}} = \frac{4}{3} \g_M \d\, .
\end{equation}
The axial couplings $g_A$ and $g_{\pi N\D}$ are well known phenomenologically.
At this order in the chiral expansion, $g_A$ can be either the nucleon axial charge or its chiral limit value,  
with the difference being of higher order than NNLO.
The quantity $\D \equiv m_\D - m_N$ is the delta-nucleon mass splitting, which is $\D\simeq293$~MeV at the physical pion mass.
$\mc{J}(m_\pi,\D,\mu)$ is a non-analytic function appearing above, defined as~\cite{Tiburzi:2005na}%
\footnote{Compared with the more standard definition of $\mc{J}^\prime$, found for example in Refs.~\cite{WalkerLoud:2004hf,Tiburzi:2004rh}, following Ref.~\cite{Tiburzi:2005na}, we define $\mc{J}(m,\D,\mu) = \mc{J}^\prime(m,\D,\mu)-\mc{J}^\prime(0,\D,\mu)$ with a suitable absorption of analytic pion mass terms in the LECs.}
\begin{multline}
\mc{J}(m,\D,\mu) = 
	2\D \sqrt{\D^2 -m^2} \ln \left( \frac{\D - \sqrt{\D^2 -m^2 + i \varepsilon}}{\D +\sqrt{\D^2 -m^2 + i \varepsilon}} \right)
\\
	+m^2 \ln \left(\frac{m^2}{\mu^2} \right)
	+2\D^2 \ln \left(\frac{4\D^2}{m^2} \right)\, .
\end{multline}
For $m>\D$, we can use the equality between $\ln$ and $\arctan$ to express this function with all positive and real arguments:
\begin{multline}
\sqrt{\D^2 -m^2} \ln \left( \frac{\D - \sqrt{\D^2 -m^2+ i \varepsilon}}{\D +\sqrt{\D^2 -m^2+ i \varepsilon}} \right)
\\
= 2 \sqrt{m^2 - \D^2} \arctan \left( \sqrt{\frac{m^2}{\D^2} -1}\right)\, .
\end{multline}
In Eq.~\eqref{eq:dMN_full}, 
the last contribution arises from the PQ effect but comes with no new LECs.
The simplification of this PQ effects occurs because of the symmetric splitting of the valence quark masses about the degenerate sea quark mass, Eq.~\eqref{eq:symm_isospin}, with the definition~\cite{WalkerLoud:2009nf}
\begin{equation}
\D_{PQ}^2 = 2B\d\, .
\end{equation}
For this choice of PQ LQCD, the same quantity which controls the isospin breaking effects also controls the PQ effects.
Lastly, $g_0$ is the singlet axial coupling which can be reliably estimated phenomenologically.

We would like to assess the various contributions to $\d M_N^\d$ arising in Eq.~\eqref{eq:dMN_full}.
At LO in $\chi$PT, $\D_{PQ}^2 = \D m_K^2$, Eq.~\eqref{eq:kaonSplitting}, so the size of the PQ corrections can be readily estimated.  Normalizing the PQ correction by the LO term, and using our computed values of $\D m_K^2$ from Table~\ref{kaonSplittingTable} as estimates for $\D_{PQ}^2$, we find
\begin{equation}
\e_{PQ} \equiv \frac{\d M_N^{\d,PQ}}{\d M_N^{\d,LO}} 
	= \frac{(4-3g_0^2) \D_{PQ}^4}{2 m_\pi^2 (4\pi f_\pi)^2}
	\lesssim 5 \cdot 10^{-4}\, ,
\end{equation}
for all values of the parameters used in this work.  
The bound is derived from the lightest pion mass, where this effect is the largest.
This is consistent with the observation that our results in Table~\ref{tab:mn_split} show no sign of quadratic $a_t \d$ dependence.
Thus, the PQ effects can be safely ignored as they are much smaller than our other uncertainties.

%
\subsection{$\chi$-extrapolation of $\d M_N^\d$\label{sec:dMN_nodelta}}
We begin with the simplest extrapolation using only the nucleon and pion degrees of freedom ,for which the quark-mass dependence is given by
\begin{multline}\label{eq:dMN_nodelta}
\delta M_{N}^{\delta} = \delta \bigg\{
	\a_N \left[1
		-\frac{m_\pi^2}{(4\pi f)^2}(6g_A^2+1)\ln \left(\frac{m_{\pi}^2}{\mu^2}\right)
	\right] 
\\
	+ \beta(\mu)\frac{2m_{_\pi}^2}{(4 \pi f)^2}\bigg\}\, .
\end{multline}
In this work, we have not computed the pion decay constant or the nucleon axial coupling.
While the pion decay constant has a relatively large pion-mass dependence, it is know that the nucleon axial coupling has a very mild pion-mass dependence.
For a recent review including $g_A$, see Ref.~\cite{collins:latt2016}.
Whether we take $f$ to be the chiral-limit value of $f_\pi$, the physical value or pion-mass dependent, the differences are all higher order than NNLO.
For our central values, we take $f=f_\pi^{phy} = 130.4$~MeV.
Because we are interested in identifying the presence of the chiral logarithm in Eq.~\eqref{eq:dMN_nodelta}, we try setting the nucleon axial coupling to its physical value $g_A=1.2723$ and also letting it float as a free parameter in the minimization.
It is worth noting that fits to the isoscalar nucleon mass, with $g_A$ left a free parameter, return values consistent with 0 or significantly smaller than the measured value~\cite{WalkerLoud:2008bp}.
This is due, in part, to the dramatic pion-mass dependence observed in LQCD spectrum calculations in which the nucleon mass scales linearly in the pion mass~\cite{WalkerLoud:2008pj,Walker-Loud:2013yua}.

\begin{figure}
\begin{tabular}{c}
\includegraphics[width=\columnwidth]{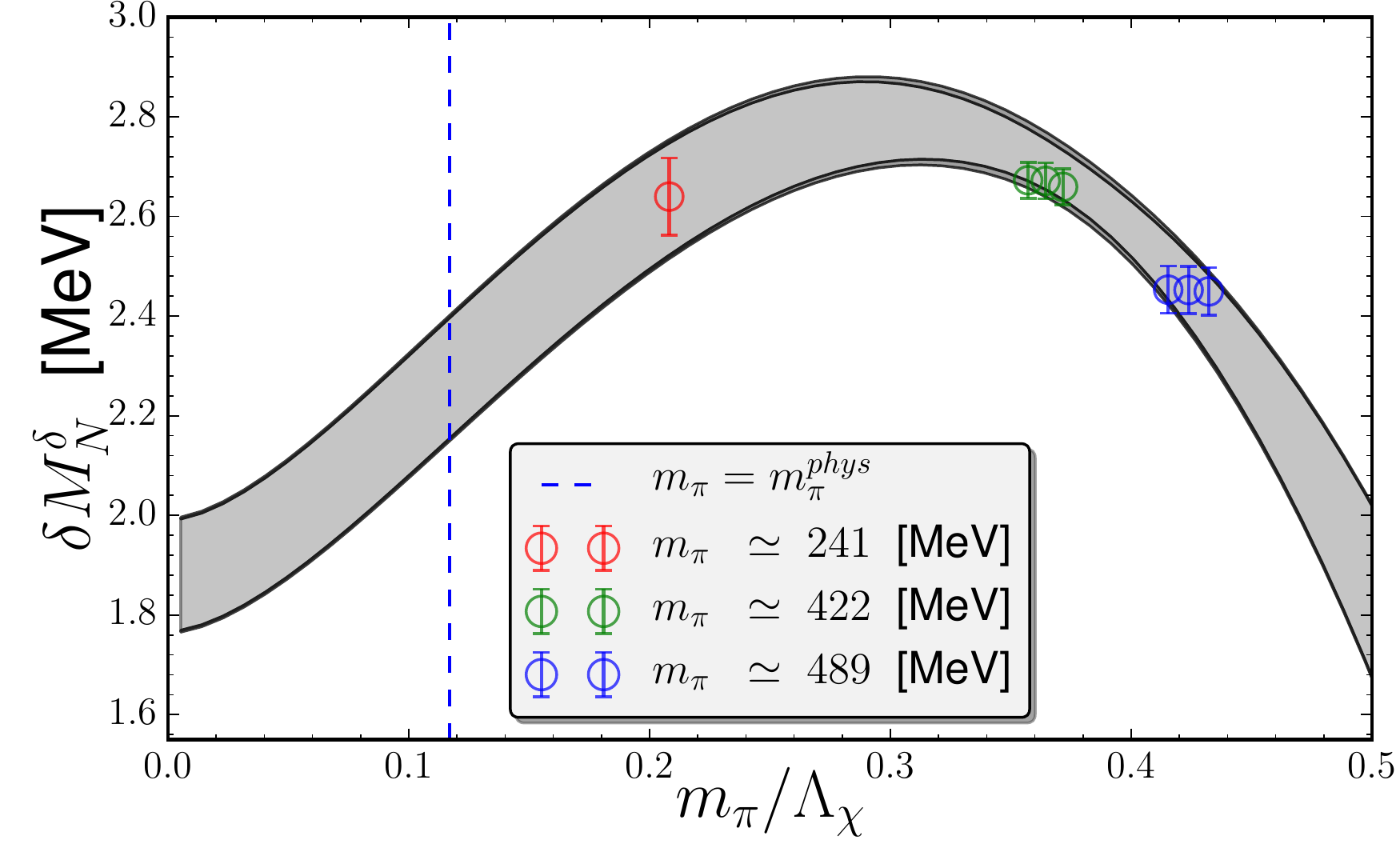}
\end{tabular}
\caption{\label{fig:dMN_nodelta} The nucleon mass splitting $\d M_N^\d$ versus $m_\pi/\L_\chi$ where $\L_\chi = 2\sqrt{2}\pi f$.  
The numerical results show statistical uncertainties only.
The multiple values at the two heavier pion masses arise from the three values of $a_t\d$ used in this work and are split for clarity.  These values have been converted to MeV and scaled to the physical quark mass splitting $a_t\d^*$, Eq.~\eqref{eq:delta}.}
\end{figure}

In the first extrapolation analysis we perform, we set $g_A = 1.2723$.  
With this value, Eq.~\eqref{eq:dMN_nodelta} predicts a strong pion-mass dependence due to the large coefficient in front of the logarithm, $6g_A^2 + 1$.
The resulting fit is tabulated in Table~\ref{tab:dMN_nodelta} and depicted in Figure~\ref{fig:dMN_nodelta}, and produces the value 
\begin{equation}\label{eq:dMN_pred}
	\d M_{N}^{\d} = 2.28(11)(3)(5) \textrm{ MeV}\, .
\end{equation}
The first uncertainty is from combined statistical and systematic uncertainties in the correlator analysis.  The second uncertainty is from the value of $a_t \d^*$ we determine, Eq.~\eqref{eq:delta}, and the third uncertainty is from our scale setting analysis, Eq.~\eqref{eq:physical_scale}.
As is evident from the quality of fit, this extrapolation is strongly favored by our numerical results.
The strong curvature arises from the competition between the logarithm and the local counter-term $\beta$ in Eq.~\eqref{eq:dMN_nodelta}.
This very rapid pion-mass dependence is precisely what cannot be accounted for easily in a power-series expansion about $m_\pi=0$, but is easily accommodated using the extrapolation formula predicted by $\chi$PT.
A detailed study of power-series expansion fits shows that the size of the higher-order terms are as large or larger than the lower-order terms, 
and the result is unstable with respect to the inclusion of higher-order terms.

\begin{table}
\begin{ruledtabular}
\begin{tabular}{ccc|cc|cl}
 $\alpha_N$ & $\beta$& $g_A$
& $\chi^2/$dof& $Q$ & $a_{t}\delta M_{N}^{\d}$~MeV
\\
\hline 
 1.64(09)& -5.2(1.3)& fixed& 2.73/5& 0.74&2.28(11)(3)(5)\\
1.67(47)& -5.1(2.3)& 1.24(56)& 2.72/4& 0.61&2.29(32)(3)(5)\\
\end{tabular}
\end{ruledtabular}
\caption{\label{tab:dMN_nodelta} 
Chiral extrapolation of $\d M_N^\d$ using Eq.~\eqref{eq:dMN_nodelta} with $g_A$ input (fixed) or free to float in the minimization.
}
\end{table}

%
\subsubsection{Support for a large $\chi$-log coefficient in the LQCD results\label{sec:gAfloat}}

From the perspective of exposing non-analytic light quark-mass dependence, the most interesting prospect in our analysis is to relax the input of $g_A$ and see what value the numerical results favor.
In the subsequent analysis, we let $g_A$ float and only input the value of $f_\pi$, 
which we take to be the physical pion decay constant, 
as above.
The resulting fit results are provided in Table~\ref{tab:dMN_nodelta}.
As demonstrated by this analysis, the numerical results strongly favor a large coefficient of the $\chi$-logarithm, with a value of nucleon axial coupling
\begin{equation}\label{eq:ga_float}
 	g_A = 1.24(56)\, .
\end{equation}
While there is a large uncertainty on the axial coupling, it is very encouraging that the numerical results for the isovector mass prefer a large value, as this is the coefficient of the $\chi$-logarithm.
This is in sharp contrast to the numerical analysis of the isoscalar nucleon mass~\cite{WalkerLoud:2008bp}, where floating $g_A$ results $g_A\lesssim 0.4$. 
This observation quantitatively justifies for the first time our choice to input the value of $g_A=1.2723$ to our analysis.

%
\subsubsection{Influence of heaviest pion mass on the $\chi$-log}
\begin{figure}
\begin{tabular}{c}
\includegraphics[width=\columnwidth]{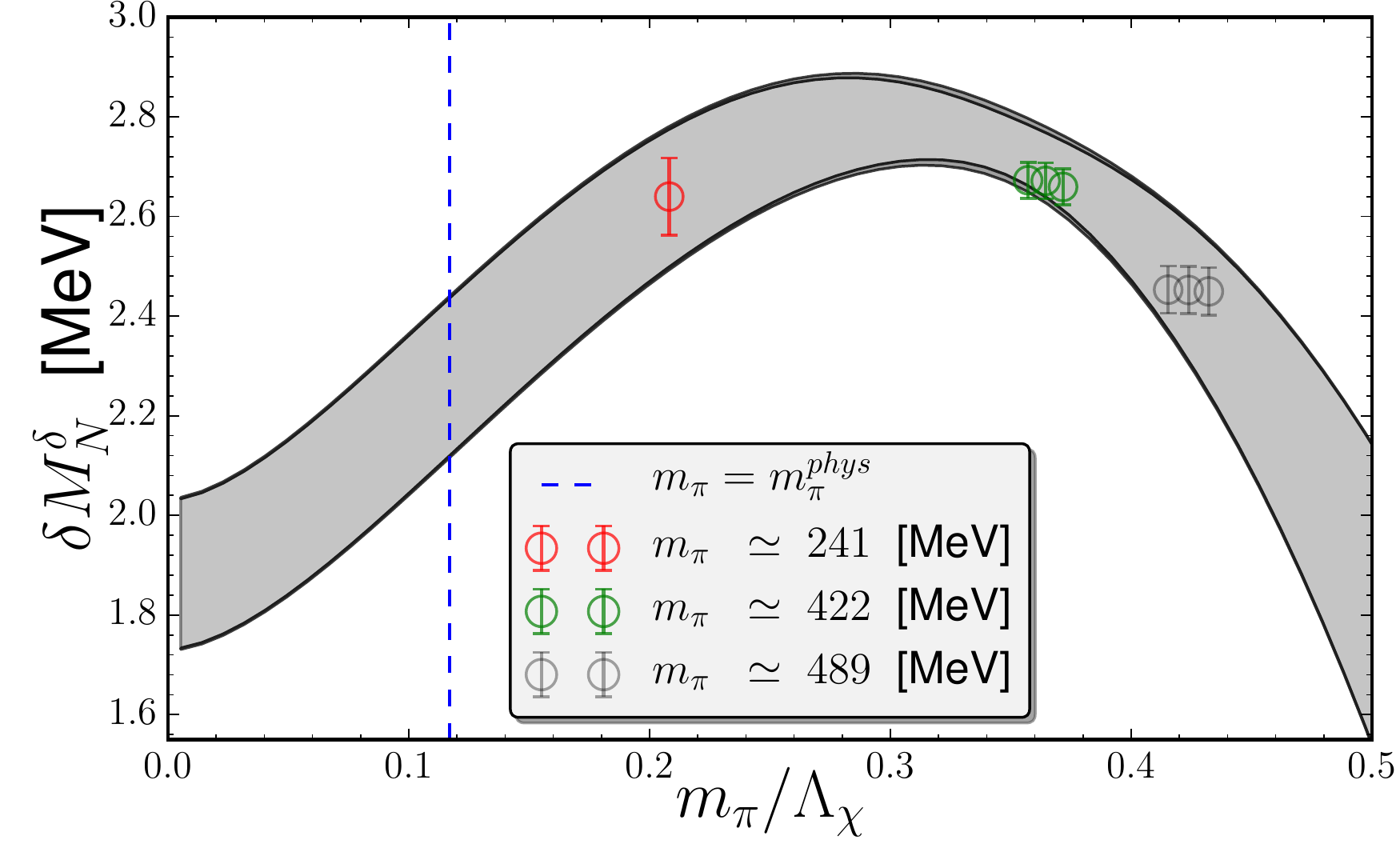}
\end{tabular}
\caption{Analysis of $\d M_N^\d$ excluding the heaviest pion mass results. 
\label{fig:dmN_cut}}
\end{figure}

One may worry that the largest pion mass data strongly influences the fit and induces the curvature.
To test this, we drop the heaviest pion mass results from the analysis, resulting in the fit depicted in Figure~\ref{fig:dmN_cut}.
As is evident, the resulting fit is in perfect agreement, but less precise, indicating the heaviest pion mass results align with the predicted $\chi$PT formula, and only serve to improve the precision of the analysis.
The resulting nucleon mass splitting in this case is $\d M_N^\d = 2.28(15)(03)(05)$~MeV, to be compared to Eq.~\eqref{eq:dMN_pred}.

%
\subsubsection{$\D$-full extrapolation\label{sec:dMN_nodelta}}
The last chiral extrapolation systematic we explore is whether the numerical results are sensitive to the delta-resonance contributions.
For $m_\pi \gtrsim 290$~MeV, the delta-resonance becomes stable as $m_N + m_\pi > m_\D$ in this pion-mass regime.
The delta degrees of freedom are also strongly coupled to the nucleon with $g_{\pi N\D} \simeq 1.5$.
For these reasons, there is an expectation that these contributions will be important to include explicitly.
Neglecting the delta degrees of freedom is equivalent to integrating them out using a small expansion parameter of $\e_\pi^\D = m_\pi / \D$, which is clearly not small for LQCD calculations with pion masses heavier than physical.

\begin{table}
\begin{ruledtabular}
\begin{tabular}{cccccc}
$a_t m_l$& $a_t m_s$& $m_\pi$& $a_t \d$& $\D $& $\D_{\Xi^* \Xi}$ \\
&&[MeV]& & [MeV]& [MeV]\\
\hline
-0.0860& -0.0743& 241& 0.0002& 330(12)(12)& 244(06)(06) \\
-0.0840& -0.0743& 421& 0.0002& 318(12)(06)& 257(06)(06) \\
-0.0840& -0.0743& 421& 0.0004& 318(12)(06)& 257(06)(06) \\
-0.0840& -0.0743& 421& 0.0010& 318(12)(06)& 263(06)(06) \\
-0.0830& -0.0743& 490& 0.0002& 244(24)(18)& 232(12)(06) \\
-0.0830& -0.0743& 490& 0.0004& 244(24)(18)& 232(12)(06) \\
-0.0830& -0.0743& 490& 0.0010& 244(24)(18)& 232(12)(06) \\
\end{tabular}
\end{ruledtabular}
\caption{\label{tab:d_n_split_xis_xi_split}
The delta-nucleon ($a_t \D$) and cascade ($a_t \D_{\Xi^* \Xi}$) mass splittings determined in this work, for different values of $a_t\d$ and $m_\pi$ on the various ensembles.}
\end{table}

In order to assess whether our numerical results support the inclusion of the delta degrees of freedom, we perform several different analyses.
In each assessment, we use the extrapolation formula 
\begin{multline}\label{eq:dMN_wdelta}
\delta M_{N}^{\delta} = \delta \bigg\{
	\a_N \left[1
		-(6g_A^2+1)\frac{m_\pi^2}{(4\pi f_{\pi})^2}\ln \left(\frac{m_{\pi}^2}{\mu^2}\right)
	\right] 
\\
	+ 4 g_{\pi N\D}^2 \left(\frac{20}{9} \g_M -\a_N \right) \frac{\mc{J}(m_\pi,\D,\mu)}{(4\pi f_\pi)^2}
\\
	+ \beta(\mu)\frac{2m_{_\pi}^2}{(4 \pi f_{\pi})^2}
	\bigg\}\, .
\end{multline}
To perform the analysis, we also determine or estimate the values of $\D=m_\D-m_N$ and $\g_M$ using the delta correlation functions.
The values of $\D$ are collected in Table~\ref{tab:d_n_split_xis_xi_split} and the isospin splittings $m_{\D^-} - m_{\D^0}$ in Table~\ref{tab:delta_xi_splittings}.
From Eq.~\eqref{eq:gamma_m}, we see only the product $\g_M \d$ is renormalization scheme and scale independent and as we are working with bare values of $\d$. 
Thus, we find estimate the unrenormalized LEC 
\begin{equation}
	\mathring{\g}_M = 1.68(3)\, .
\end{equation}
Given the similarity of our estimate of the condensate $\mathring{\S}$, Eq.~\eqref{eq:bareSigma} with that in the FLAG report~\cite{Aoki:2016frl}, we expect this to be a good approximation of the renormalized LEC.

\begin{table}
\begin{ruledtabular}
\begin{tabular}{cccccc}
$a_t m_l$& $a_t m_s$& $m_\pi$& $a_t \d$& $m_{\D^{-}} - m_{\D^{0}} $ & $m_{\Xi^{*-}} - m_{\Xi^{*0}} $ \\
&&[MeV]& & [MeV]& [MeV] \\
\hline
-0.0860& -0.0743& 241& 0.0002 & -- & 3.09(14)(24)\\
-0.0840& -0.0743& 421& 0.0002 & 2.80(05)(12) & 2.86(04)(08)\\
-0.0840& -0.0743& 421& 0.0004 & 5.56(10)(24) & 5.72(09)(15)\\
-0.0840& -0.0743& 421& 0.0010 & 13.6(23)(55) & 14.2(02)(04)\\
-0.0830& -0.0743& 490& 0.0002 & 2.52(06)(08) & 2.68(06)(12)\\
-0.0830& -0.0743& 490& 0.0004 & 5.05(13)(15) & 5.36(12)(24)\\
-0.0830& -0.0743& 490& 0.0010 & 12.7(03)(04) & 13.3(03)(05)\\
\end{tabular}
\end{ruledtabular}
\caption{\label{tab:delta_xi_splittings}
The $\D$ baryon mass splitting used in the determination of $\gamma_{m}$ e.g. Eq.~\eqref{eq:gamma_m}. The $\Xi^*$ baryon mass splitting is used to determine $\alpha_{\Xi^*}$ as in  Eq.~\eqref{eq:alpha_xis}. As the $\D$ baryon is unstable at the lightest pion mass, no fit was taken from this ensemble. }
\end{table}

When assessing the contribution of these new terms, we always take $g_A = 1.2723$ because that is consistent with our unrestricted analysis in Sec.~\ref{sec:gAfloat}.
The leading large-$N_c$ relation between $g_A$ and $g_{\pi N\D}$ provides the estimate~\cite{Dashen:1994qi,Jenkins:1995gc}
\begin{equation} \label{eq:prior_gDN}
g_{\pi N\D} = \frac{6}{5} g_A + \mc{O}\left(\frac{1}{N_c}\right)\, .
\end{equation}
We perform the analysis of our results using Eq.~\eqref{eq:dMN_wdelta} augmented with Bayesian constrained fits with several generous values of a Gaussian prior width.%
\footnote{There has been a recent interest in using Bayesian analysis methods for determining LECs in EFTs~\cite{Schindler:2008fh,Furnstahl:2014xsa,Furnstahl:2015rha,Wesolowski:2015fqa}.} 
The results are collected in Table~\ref{tbl:dMN_wDelta}.
All fits have a good fit-statistic and the predicted values of $\d M_N^\d$ are largely insensitive to these modifications.
However, we observe that the extracted uncertainty on the $g_{\pi N\D}$ axial coupling tracks the size of the prior width indicating the numerical results provide no guidance for the delta contributions.
The strongest conclusion one can draw from this analysis is that the numerical results are not inconsistent with the contributions from the delta degrees of freedom, but there is no quantitative support for them.

\begin{table}
\begin{ruledtabular}
\begin{tabular}{ccccc}
 $\tilde{g}_{\pi N\D}$& $\alpha_N$ & $\beta$& $\hat{g}_{\pi N\D}$ & $\delta M_{N}^{\d}$~MeV
\\
\hline 
1.50(25)& 1.79(10)& -16(3)& 1.51(25)& 2.40(12)(4)(5)\\
1.50(50)& 1.78(12)& -15(6)& 1.46(47)& 2.39(14)(4)(5)\\
1.50($\infty$)& 1.66(32)& -6(22)& 0.49(4.47)& 2.29(28)(4)(5)\\
\end{tabular}
\end{ruledtabular}
\caption{\label{tbl:dMN_wDelta} 
Chiral extrapolation of $\d M_N^\d$ using Eq.~\eqref{eq:dMN_wdelta} with a Bayesian constraint on $g_{\pi N\D}$.  The prior width given to the augmented $\chi^2$ is denoted $\tilde{g}_{\pi N\D}$ and $\hat{g}_{\pi N\D}$ is the resulting posterior value.
For any small finite prior width, the coupling is just determined by the prior, 
Eq~\eqref{eq:prior_gDN}.
}
\end{table}

%
\subsubsection{$\d M_\Xi^\d$ and the lack of $\chi$-logarithmic behavior}
The cascade also forms an isodoublet, like the nucleon.
At low-energies, the $SU(2)$ $\chi$PT theory for the $\Xi$ will be identical in form to that of the nucleon with only numerical values of the LECs being different, as reflected in Eqs.~\eqref{eq:LO_N} and \eqref{eq:LO_Xi}.
Including virtual corrections from the resonant spin-3/2 $\Xi^*$ states breaks the exact mapping of Eq.~\eqref{eq:dMN_full} to the $\Xi,\Xi^*$ system, 
as the $\Xi^*$ form an iso-doublet while the $\D$ states form an iso-quartet.
Accounting for these differences, the full expression for the iso-vector $\Xi$ mass becomes
\begin{multline}\label{eq:dMXi_wdelta}
\delta M_{\Xi}^{\delta} = \delta \bigg\{
	\a_\Xi \left[1
		-(6g_{\pi\Xi\Xi}^2+1)\frac{m_\pi^2}{(4\pi f_{\pi})^2}\ln \left(\frac{m_{\pi}^2}{\mu^2}\right)
	\right] 
\\
	+ g_{\pi\Xi\Xi^*}^2 \left(4 \a_{\Xi^*} -3\a_\Xi \right) \frac{\mc{J}(m_\pi,\D_{\Xi^*\Xi},\mu)}{(4\pi f_\pi)^2}
\\
	+ \beta_\Xi(\mu)\frac{2m_{_\pi}^2}{(4 \pi f_{\pi})^2}
	\bigg\}\, .
\end{multline}
This expression can be determined from Ref.~\cite{deVries:2015una} by matching $SU(3)$ onto $SU(2)$ $\chi$PT~\cite{Tiburzi:2008bk}. We use the LO contribution to the $\Xi^*$ isospin splitting to determine $\alpha_{\Xi^*}$, e.g. 
\begin{equation}\label{eq:alpha_xis}
m_{\Xi^{*-}} -  m_{\Xi^{*0}} = -4 \alpha_{\Xi^*} \d,
\end{equation}
with the data collected in Table~\ref{tab:delta_xi_splittings}.
This allows an estimation of the unrenormalized LEC
\begin{equation}
	\mathring{\a}_{\Xi^*} = -0.58(2)\, .
\end{equation}

\begin{table*}
\begin{ruledtabular}
\begin{tabular}{cc|ccccc}
 $\tilde{g}_{\pi \Xi \Xi}$&$\tilde{g}_{\pi \Xi \Xi^*}$&$\hat{g}_{\pi \Xi \Xi}$&  $\hat{g}_{\pi \Xi \Xi^*}$ &$\alpha_{\Xi}$ & $\beta_{\Xi}$ & $\delta M_{\Xi}^{\d}$~MeV
\\
\hline
 \multicolumn{7}{c}{\rule{0pt}{2.5ex}$\Delta_{\Xi \Xi^*} =213.5$~MeV}\\
\hline
0.240(02)&0.882(09)& 0.240(02)& 0.882(09)& 4.59(22)& -2.6(14)& 5.37(24)(8)(5)\\
0.240(05)&0.882(18) &0.240(05)& 0.882(18)& 4.59(23)& -2.6(15)& 5.37(24)(8)(5)\\
0.240(12)&0.882(44)&0.240(12)& 0.882(44)& 4.59(22)& -2.6(15)& 5.37(24)(8)(5)\\
0.240(24)&0.882(88)&0.240(24)& 0.885(88)& 4.59(23)& -2.6(19)& 5.37(25)(8)(5)\\
\hline
 \multicolumn{7}{c}{\rule{0pt}{2.5ex}$\Delta_{\Xi \Xi^*}  = \Delta_{\Xi \Xi^*}^{LQCD}$~MeV}\\
\hline
0.240(02)&0.882(09)& 0.240(02)& 0.882(09)& 4.70(24)& -2.3(15)& 5.50(25)(8)(5)\\
0.240(05)&0.882(18) &0.240(05)& 0.882(18)& 4.70(24)& -2.3(16)& 5.50(25)(8)(5)\\
0.240(12)&0.882(44)&0.240(12)& 0.882(44)& 4.70(24)& -2.3(17)& 5.50(25)(8)(5)\\
0.240(24)&0.882(88)&0.240(24)& 0.884(88)& 4.70(24)& -2.3(20)& 5.50(26)(8)(5)\\

\end{tabular}
\end{ruledtabular}
\caption{\label{tbl:dMXI_priorAxial} 
Chiral extrapolation of $\d M_{\Xi}^\d$ using Eq.~\eqref{eq:dMXi_wdelta} with Bayesian constrained fits.  The prior values are denoted as $\tilde{g}_{\pi \Xi \Xi}$ while the posteriors are denoted as $\hat{g}_{\pi \Xi \Xi}$.
}
\end{table*}

\begin{figure}
\begin{tabular}{c}
\includegraphics[width=\columnwidth]{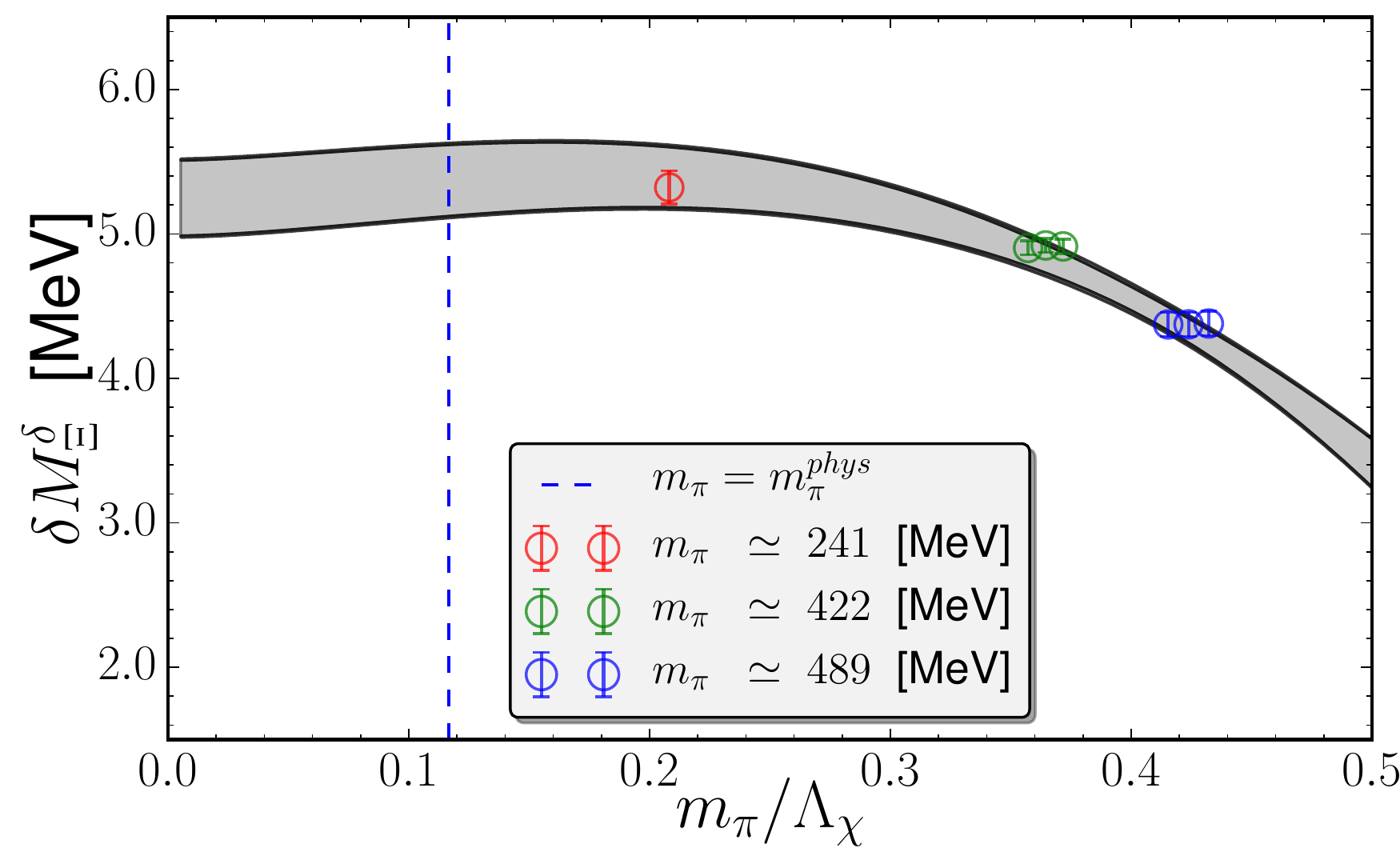}
\end{tabular}
\caption{\label{fig:dMXi} The mass splitting $\d M_\Xi^\d$ [MeV] versus $m_\pi$ with $g_{\pi\Xi\Xi}$ and $g_{\pi\Xi\Xi^*}$ constrained with 5\% prior widths.  
The numerical results show statistical uncertainties only.
The multiple values at the two heavier pion masses arise from the three values of $a_t\d$ used in this work and are split for visual clarity.}
\end{figure}

Phenomenologically, we know the $\Xi$ axial charge is much smaller than the nucleon axial charge.  Similarly, the axial transition coupling is suppressed~\cite{Tiburzi:2008bk},
\begin{align} \label{eq:priorXi}
&g_{\pi\Xi\Xi} \simeq 0.24\, ,&
&g_{\pi\Xi\Xi^*} \simeq \frac{g_{\pi N\D}}{\sqrt{3}} \simeq 0.87\, .&
\end{align}
Comparing the coefficient of the chiral-log term arising from the $\Xi-\pi$ virtual state, we estimate that this logarithmic $m_\pi$ contribution is $\mc{O}(10)$ times smaller than in $\d M_N^\d$.
We observe the pion-mass dependence of $\d M_{\Xi}^\d$ is much milder than that of the nucleon, see Figure~\ref{fig:dMXi}.
However, the contribution from the $\Xi^*-\pi$ virtual corrections is not as suppressed.

In order to assess the contributions from the $\Xi^*$ states, we therefore perform an analysis using Bayesian priors on both axial couplings, $g_{\pi\Xi\Xi}$ and $g_{\pi\Xi\Xi^*}$.
We explore setting prior widths that are 1, 2, 5 and 10\% of the phenomenological values in Eq.~\eqref{eq:priorXi}.
We use both the experimental $\D_{\Xi^*\Xi}$ splitting as well as those determined in this work, see Table~\ref{tab:d_n_split_xis_xi_split}.
The results of these analyses are collected in Table~\ref{tbl:dMXI_priorAxial} and a representative extrapolation is depicted in Figure~\ref{fig:dMXi}.
As with the nucleon isovector mass, we find the uncertainty on $g_{\pi\Xi\Xi}$ and $g_{\pi\Xi\Xi^*}$ scales with the prior width we set.  However, we also observe the resulting value of $\d M_\Xi^\d$ is stable as we increase the prior width.
We therefore take the results with 5\% prior widths on the axial couplings.
There is a systematic associated with using the experimental value of $\D_{\Xi^*\Xi}$ and the values determined in this work, which is nominally higher order.  
For our final prediction, we therefore split this difference as a systematic
\begin{equation}\label{eq:dMXi_final}
\d M_\Xi^\d = 5.44(24)(8)(5)(7) \textrm{ MeV}\, ,
\end{equation}
where the uncertainties are the fitting statistical/systematic uncertainty, the uncertainty from $a_t \d^*$, the scale-setting uncertainty and finally the uncertainty from $\D_{\Xi^*\Xi}$.

%
\subsubsection{$\chi$-logarithms in the isovector nucleon mass}
Taken all together, we find the evidence presented here to be conclusive evidence for the presence of non-analytic light quark mass dependence in the nucleon spectrum:
\begin{itemize}
\item strong pion-mass dependence is observed which cannot be accounted for with a power-series expansion about $m_\pi=0$, Figure~\ref{fig:dMN_nodelta} but perfectly predicted and accounted for with $\chi$PT;

\item the observed pion-mass dependent curvature is not sensitive to the inclusion of the heaviest pion mass data, Figure~\ref{fig:dmN_cut};

\item relaxing the coefficient of the $\chi$-log to freely vary results in the large value of $g_A$ consistent with the experimental value, Eq.~\eqref{eq:ga_float};

\item the lack of observation of strong pion-mass dependence in the cascade isovector mass, which is in accordance with expectations predicted by $\chi$PT, Figure~\ref{fig:dmN_cut}.

\end{itemize}

\begin{figure}
\begin{tabular}{c}
\includegraphics[width=\columnwidth]{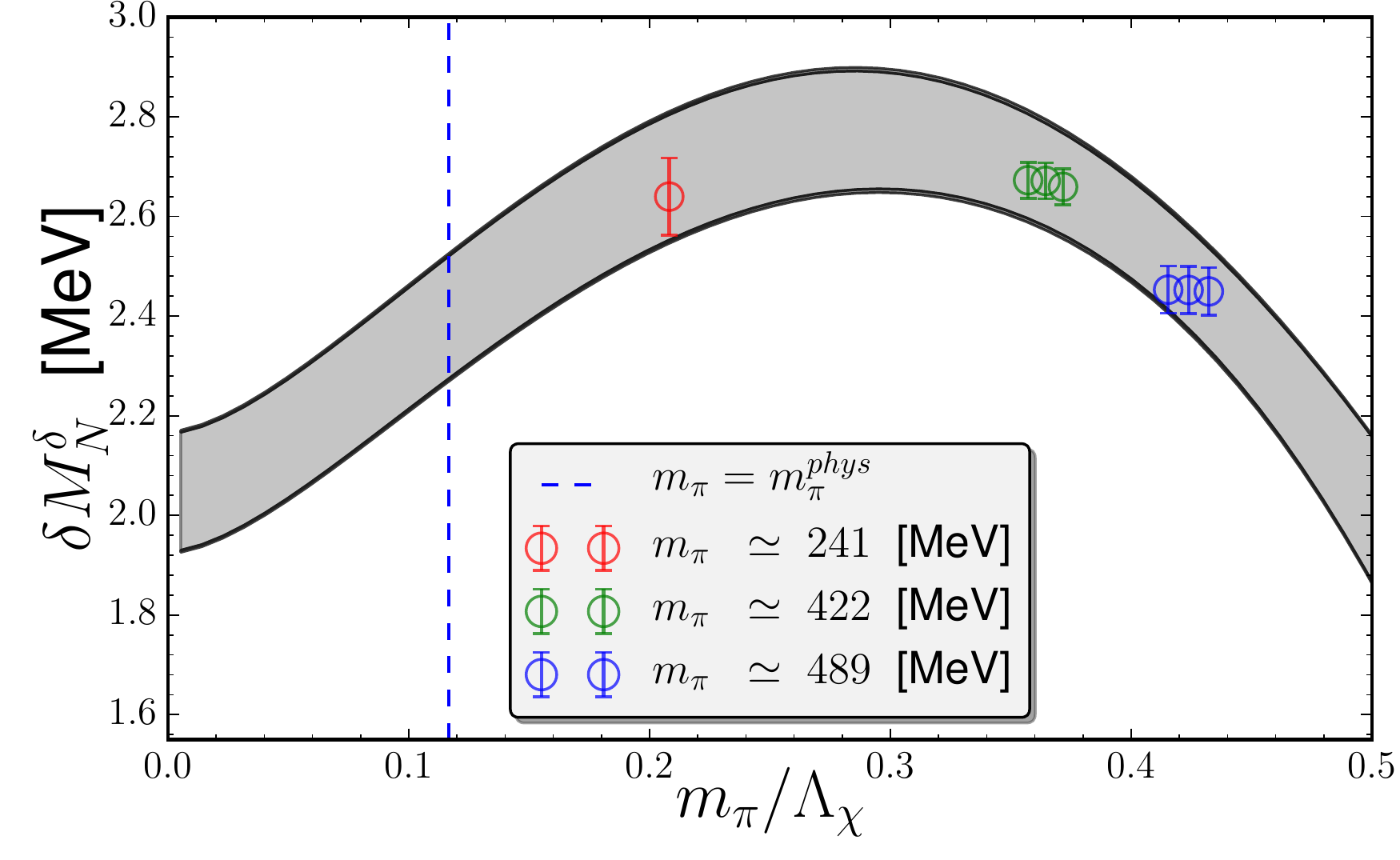}
\end{tabular}
\caption{\label{fig:dMN_final}
The mass splitting $\d M_N^\d$ [MeV] versus $m_\pi$ with $g_{A}$ and $g_{\pi N\D}$ constrained with 5\% prior widths.  
The numerical results show statistical uncertainties only.
The multiple values at the two heavier pion masses arise from the three values of $a_t\d$ used in this work and are split for visual clarity.}
\end{figure}

To be conservative, for our final determination of $\d M_N^\d$, we use a fit including both nucleon and delta intermediate states, Eq.~\eqref{eq:dMN_wdelta}.
We use our prior knowledge of $g_A$ and $g_{\pi N\D}$ from experiment to allow these couplings to float in the numerical minimization, but we apply reasonable prior widths to their central values via an augmented $\chi^2$ with Gaussian priors.
We explore the sensitivity of the extrapolated value of $\d M_N^\d$ to the size of the prior widths on these axial couplings with 1, 2, 5 and 10\% widths.
As with $\d M_\Xi^\d$, we use both the value of $\D = m_\D-m_N$ from experiment, and determined in this work, Table~\ref{tab:d_n_split_xis_xi_split} as a further extrapolation systematic.
The results of this study are presented in Table~\ref{tbl:dMN_priorAxial}.
The posterior uncertainties on the axial couplings track the prior widths, however, the resulting value of $\d_M^\d$ is not sensitive to this variation.
We observe dependence on the values of $\D$, which we take as an extrapolation systematic.
Our final prediction for the strong contribution to the isovector nucleon mass is 
\begin{align}\label{eq:dm_pred}
	\d M_N^\d &= 
	2.32(12)(4)(5)(8) \textrm{ MeV}\, ,
\end{align}
where the uncertainties are the fitting statistical/systematic uncertainty, the uncertainty from $a_t \d^*$, the scale-setting uncertainty and finally the uncertainty from $\D$.
A representative fit is provided in Figure~\ref{fig:dMN_final}.

\begin{table*}
\begin{ruledtabular}
\begin{tabular}{cc|ccccc}
 $\tilde{g}_{A}$&$\tilde{g}_{\pi N\D}$&$\hat{g}_{A}$&  $\hat{g}_{\pi N\D}$ &$\alpha_N$ & $\beta$ & $\delta M_{N}^{\d}$~MeV
\\
\hline
 \multicolumn{7}{c}{\rule{0pt}{2.5ex}$\Delta =293$~MeV}\\
\hline
1.27(01)&1.53(02)& 1.27(01)& 1.53(02)& 1.80(09)& -15.9(08)& 2.40(11)(4)(5)\\
1.27(03)&1.53(03) &1.27(03)& 1.53(03)& 1.80(09)& -15.9(09)& 2.40(12)(4)(5)\\
1.27(06)&1.53(08)&1.28(06)& 1.53(08)& 1.80(10)& -16.0(14)& 2.40(12)(4)(5)\\
1.27(13)&1.53(15)&1.29(12)& 1.52(15)& 1.79(14)& -16.0(24)& 2.40(13)(4)(5)\\
\hline
 \multicolumn{7}{c}{\rule{0pt}{2.5ex}$\Delta = \D_{LQCD}$~MeV}\\
\hline
1.27(01)&1.53(02)& 1.27(01)& 1.53(02)& 1.67(12)& -15.2(12)&  2.23(14)(3)(5)\\
1.27(03)&1.53(03) &1.27(03)& 1.53(03)& 1.67(12)& -15.2(12)& 2.23(12)(3)(5)\\
1.27(06)&1.53(08)&1.26(06)& 1.51(08)&  1.67(13)& -15.0(16)& 2.23(12)(3)(5)\\
1.27(13)&1.53(15)&1.24(13)& 1.48(15)& 1.68(15)& -14.2(26)& 2.24(15)(3)(5)\\

\end{tabular}
\end{ruledtabular}
\caption{\label{tbl:dMN_priorAxial} 
Chiral extrapolation of $\d M_N^\d$ using Eq.~\eqref{eq:dMN_wdelta} with Bayesian constrained fits.  Here $\tilde{g}_{\pi N\D}$ is the prior width given to the augmented $\chi^2$ and $\hat{g}_{\pi N\D}$ is the resulting fit value.
For any small finite prior width, the coupling is just determined by the prior, 
Eq~\eqref{eq:prior_gDN}.
}
\end{table*}

%% file: sections/4_theta.tex
\section{Implications for the QCD $\theta$-term\label{sec:theta}}

CP (Charge-Parity) violation from the QCD $\bar\theta$ term is intimately related to the quark masses \cite{tHooft:1976rip,Baluni:1978rf,Crewther:1979pi}. 
Via the $U(1)_A$ anomaly, the $\bar\theta$ term can be rotated into a complex quark-mass term, which, 
after performing additional non-anomalous $SU(N_f)$ rotations needed to align the vacuum of the theories with and without CP violation,
is isoscalar and proportional to the light quark reduced mass $m_*$. In SU(2),  $m_* = (1/m_u  + 1/m_d)^{-1}$ and the quark-mass operator can be expressed as
\begin{align}\label{QCDmass}
\mathcal L_m &=  
	- \bar m \bar q q + \delta\, \bar q \tau_3 q +   m_* \bar\theta \bar q i \gamma_5 q 
\nonumber\\ &= 
	- \bar m \bar q q + \delta \bar q \tau_3 q +   \frac{\bar m}{2} \left(1 - \frac{\delta^2}{\bar m^2}\right) \bar\theta \bar q i \gamma_5 q,
\end{align}
where $2\bar m = m_u + m_d$. 
The observation of Ref. \cite{Crewther:1979pi} is that the QCD $\bar\theta$ term and the quark mass difference are related by an $SU(2)_L \times SU(2)_R$ rotation, and this implies 
that chiral symmetry relates the matrix elements of the isoscalar $\bar\theta$  term between $n_N$ nucleon and $n_\pi$ pions to those
of the isovector quark-mass term with $n_N$ nucleons and $n_\pi -1$ pions.
These relations are particularly robust for the leading interactions induced by $\mathcal L_m$ in the $\chi$PT Lagrangian \cite{Mereghetti:2010tp,deVries:2015una}.
 
The pseudoscalar mass term in Eq. \eqref{QCDmass} induces isospin invariant, TV (time-reversal violating) pion-baryon couplings, 
\begin{equation}
\mathcal L = - \frac{\bar g_0}{\sqrt{2} f_\pi} \bar N \tau \cdot \pi N  - \frac{\bar g_{0\, \Xi}}{\sqrt{2} f_\pi}  \bar \Xi\, \tau \cdot \pi \Xi + \ldots, 
\end{equation}
where $\ldots$ includes terms with multiple pions, which are fixed by chiral symmetry, and TV couplings of the $\Sigma$ and $\Lambda$, which we will not discuss.
The coupling of greatest phenomenological interest is $\bar g_0$, which determines the leading non-analytic contributions to the nucleon EDM (electric dipole moment) \cite{Crewther:1979pi} and the momentum dependence of the 
nucleon EDFF (electric dipole form factor) \cite{Hockings:2005cn,Mereghetti:2010kp}. Furthermore, $\bar g_0$ dominates the nucleon-nucleon TV potential induced by the QCD $\bar\theta$ term,
and, consequently, the $\bar\theta$ term contribution to the EDM of $^{3}$He, and of diamagnetic atoms, such as $^{199}$Hg and $^{129}$Xe.

Chiral symmetry implies that, for CP violation induced by the QCD $\bar\theta$ term, the nonperturbative information entering $\bar g_0$ and $\bar g_{0\Xi}$ is determined  by the quark mass contribution to nucleon and cascade mass splittings  \cite{Crewther:1979pi,Mereghetti:2010tp}
\begin{align}
\bar g_0(\bar\theta) &= 
	\delta M_{N}^{\d} \frac{\bar m }{2 \delta} \left(1 - \frac{\delta^2}{\bar m^2}\right)  \bar\theta\, ,
\nonumber\\
\bar g_{0\, \Xi}(\bar\theta)   &= 
	\delta M_{\Xi}^{\d} \frac{\bar m }{2 \delta} \left(1 - \frac{\delta^2}{\bar m^2}\right)  \bar\theta. \label{relations}
\end{align}
These relations were derived at LO in $\chi$PT, but it has been showed that they are respected by all loop corrections of $\mathcal O(\varepsilon_\pi)$, and violated only by finite counterterms  \cite{deVries:2015una}.

Our extraction of the nucleon and cascade mass splittings allows for a precise determination of $\bar g_0$ and $\bar g_{0\Xi}$. We find  
\begin{eqnarray}
\frac{\bar g_0}{\sqrt{2} f_\pi} &=&  \left( 14.7 \pm 1.8 \pm 1.4 \right) \cdot 10^{-3}  \, \bar\theta,  \label{g0pi}\\ 
\frac{\bar g_{0\Xi}}{\sqrt{2} f_\pi} &=& \left( 34.4 \pm 4.0 \pm 3.5 \right) \cdot 10^{-3}  \, \bar\theta  \label{g0Xi}, \qquad 
\end{eqnarray}
where we used the FLAG averages for $\delta / \bar m$ at the physical point, $\delta / \bar m = 0.37\pm 0.03$ \cite{Aoki:2016frl}.
The first error in Eqs. \eqref{g0pi} and \eqref{g0Xi} comes from the lattice errors on the mass splittings and $\delta / \bar m$, combined in quadrature. The second error is an estimate of the
$\mathcal O(\varepsilon_\pi)$ corrections to Eq. \eqref{relations}, which, following Ref. \cite{deVries:2015una}, we conservatively estimate to be at the $10\%$ level.

The pion-nucleon coupling $\bar g_0$ determines the  non-analytic dependence of the neutron EDM on the pion mass \cite{Crewther:1979pi}. At NLO in $\chi$PT \cite{Crewther:1979pi,Mereghetti:2010kp,Guo:2012vf}
\begin{equation}\label{nucleonEDM}
d_n   = \bar{d}_n(\mu) +  \frac{e g_A \bar g_0}{8\pi^2 f_\pi^2} \left( \log \left(\frac{\mu^2}{m^2_\pi}\right) - \frac{\pi m_\pi}{2 m_N} \right) 
\end{equation}
where $\bar{d}_n(\mu)$ is a counterterm needed to absorb the scale dependence of the chiral loop, and a very similar expression holds for the EDM of the $\Xi$ baryon. 
Recent LQCD calculations  of the nucleon EDM induced by the QCD $\bar\theta$  term \cite{Shintani:2015vsx,Shindler:2015aqa,Guo:2015tla}  
do not yet show evidence of this non-analytic behavior. As the precision improves  and calculations at pion masses closer to the physical point are performed,
it will be important for LQCD to confirm, or, maybe more interestingly, refute the behavior predicted by Eq. \eqref{nucleonEDM}.

In addition, our calculation predicts the slope of the nucleon EDFF. Defining $S_B = - d F_B(\vec q^{\, 2})/d \vec q^{\,2}$, where $F_B$ is the EDFF of the baryon $B$  and 
$\vec q$ indicates the photon three-momentum, at the physical pion mass we find 
\begin{eqnarray}
S_{n} &=&  ( 0.69 \pm 0.08 ) \cdot 10^{-4} \, \bar\theta \, e\, \textrm{fm}^3   \\
\frac{g_{\pi\Xi\Xi} S_n}{ g_A S_{\Xi^-}} &=& \frac{\delta M_{N}^{\d}}{\delta M_{\Xi}^{\d} } \frac{ 1 - \frac{5 \pi m_\pi}{4 m_N}}{ 1 - \frac{5 \pi m_\pi}{4 m_\Xi}} = 0.30 \pm 0.02,
\end{eqnarray}
where we used the NLO $\chi$PT expression of the EDFF \cite{Mereghetti:2010tp,Guo:2012vf}.
While these predictions are of little phenomenological interest, 
since there are no plans to measure the momentum dependence of the nucleon or $\Xi$ EDFF, they provide  important benchmarks to check the validity of 
current and future LQCD calculations of baryonic EDMs.

%% file: sections/5_conclusions.tex
\section{Conclusions \label{sec:conclusions}}

We perform precise lattice QCD calculations of the ground state isovector spectrum by utilizing a symmetric breaking of isospin in the valence sector about the degenerate sea-quark mass.  
These results demonstrate the first conclusive evidence for non-analytic light-quark mass dependence in the baryon spectrum.
The quantity which prominently displays this non-analytic behavior is the isovector nucleon mass splitting.
The evidence includes the observation of rapidly changing pion mass dependence in this quantity, which cannot be simply understood with a well behave power-series expansion about the chiral limit.
The presence of the non-analytic $\chi$-log is robust to several systematic variations, including letting the coefficient of the $\chi$-log float as a free parameter.
We also observe the isovector $\Xi$ spectrum has a milder pion mass dependence, lending significant confidence in our understanding low-energy QCD through the confirmation of non-analytic pion-mass dependence predicted by $\chi$PT.

There are just a few LQCD calculations of $m_n-m_p$ in the literature~\cite{Beane:2006fk,Blum:2010ym,Horsley:2012fw,deDivitiis:2013xla,Borsanyi:2013lga,Borsanyi:2014jba,Horsley:2015eaa}, including a preliminary version of this research~\cite{WalkerLoud:2010qq}.
We provide the most precise result, although only Refs.~\cite{Borsanyi:2013lga,Borsanyi:2014jba} have complete control of all lattice systematics, notably a continuum limit.
Precise knowledge of the QCD contribution to $m_n-m_p$ will also allow for a more precise determination of the QED contribution than presently exists~\cite{Gasser:1974wd,WalkerLoud:2012bg,Gasser:2015dwa} when combined with the experimentally measured splitting.
While we are not able to perform the continuum limit of our results, we estimate the discretization effects to be $0.07$~MeV for $\d M_N^\d$ and $0.16$~MeV for $\d M_\Xi^\d$ with the assumption of either $\mc{O}(a_s^2)$ of $\mc{O}(\alpha^2 a_2)$ contributions.